\documentclass[amsmath,10t,twocolumn,superscriptaddress,footinbib,prx]{revtex4-2}

\pdfoutput=1

\usepackage{amsmath,amssymb,amsbsy,amsfonts,amsthm,bbm,bm}
\usepackage{color}
\usepackage{physics}
\usepackage{empheq}
\usepackage[colorlinks=true,allcolors=blue]{hyperref}
\usepackage[normalem]{ulem}
\usepackage{tikz}
\usetikzlibrary{calc,arrows.meta,automata,chains,positioning,shapes,matrix}
\usetikzlibrary{decorations.markings}
\usetikzlibrary{arrows, decorations.markings}
\usetikzlibrary{shapes.geometric}
\usepackage{mhchem}
\usepackage[caption=false]{subfig}
\usepackage{mathtools}
\usepackage{yfonts}
\usepackage{mathrsfs}

\definecolor{S_Blue}{RGB}{0,135,252}
\definecolor{S_Red}{RGB}{214,13,63}

\newcommand{\mA}{\mathcal{A}}
\newcommand{\mB}{\mathcal{B}}

\newcommand{\bfn}{\mathbf{n}}

\newcommand{\proj}{\hat{\mathcal{P}}}

\newcommand\bigzero{\makebox(0,0){\text{\Huge0}}}

\begin{document}

\title{Quantifying Rare Events in Stochastic Reaction-Diffusion Dynamics Using Tensor Networks}
\date{\today}

\author{Schuyler~B.~Nicholson}
\affiliation{Department of Chemistry,\
Northwestern University,\
2145 Sheridan Road, Evanston, \
Illinois 60208, USA
}

\author{Todd R. Gingrich}
\affiliation{Department of Chemistry,\
Northwestern University,\
2145 Sheridan Road, Evanston, \
Illinois 60208, USA
}
\begin{abstract}
  The interplay between stochastic chemical reactions and diffusion can generate rich spatio-temporal patterns.
  While the timescale for individual reaction or diffusion events may be very fast, the timescales for organization can be much longer.
  That separation of timescales makes it particularly challenging to anticipate how the rapid microscopic dynamics gives rise to macroscopic rates in the non-equilibrium dynamics of many reacting and diffusing chemical species.
  Within the regime of stochastic fluctuations, the standard approach is to employ Monte Carlo sampling to simulate realizations of random trajectories.
  Here, we present an alternative numerically tractable approach to extract macroscopic rates from the full ensemble evolution of many-body reaction diffusion problems.
  The approach leverages the Doi-Peliti second-quantized representation of reaction-diffusion master equations along with compression and evolution algorithms from tensor networks.
  By focusing on a Schl\"{o}gl model with one-dimensional diffusion between $L$ otherwise well-mixed sites, we illustrate the potential of the tensor network approach to compute rates from many-body systems, here with approximately $3\times 10^{15}$ microstates.
  Specifically, we compute the rate for switching between metastable macrostates, with the expense for computing those rates growing subexponentially in $L$.
  Because we directly work with ensemble evolutions, we crucially bypass many of the difficulties encountered by rare event sampling techniques\textemdash detailed balance and reaction coordinates are not needed.
\end{abstract}

\maketitle

\section{Introduction}

Connecting microscopic kinetics to emergent rates has a long history for equilibrium and nonequilibrium processes.
Complex systems often settle into metastable basins with rare transitions regulating the rates of switching between the emergent states.
In equilibrium, transition state theory~\cite{eyring1935activated, laidler1983development}, variational transition state theory~\cite{truhlar1980variational}, Grote-Hynes theory~\cite{grote1980stable,hynes1985chemical}, and transition path sampling (TPS)~\cite{dellago1998transition,bolhuis2002transition} have given strategies to estimate the transition rate.
Away from equilibrium, theories built upon a free energy landscape lose their applicability, yet trajectory sampling approaches can extract rates from microscopic dynamics, through noise-guided TPS approach~\cite{crooks2001efficient, gingrich2015preserving} or forward flux sampling (FFS)~\cite{allen2005sampling,allen2006simulating,allen2009forward}.
For all their merits, sampling approaches bring intrinsic challenges.
How many samples are required to estimate the rate? Is a good approximate reaction coordinate needed, and if so, how is it found \cite{ma2005automatic}?
How does the expense grow with larger or more complicated microscopic systems?
Motivated by those ubiquitous concerns, we set out to consider a different approach\textemdash one that effectively samples all possible trajectories as an ensemble evolution instead of propagating individual realizations.
Such a proposition may appear absurd given the exploding state space presented by the many-body problem, but we show that for the class of discrete state, continuous-time, reaction-diffusion master equations (RDME), we can evolve the full distribution of a classical many-body systems in a numerically controllable manner by combining two well-developed methods, the Doi-Peliti (DP) formalism with tensor network (TN) algorithms.

While TPS and FFS have been developed for a variety of dynamical equations of motion (Hamiltonian, Langevin, Markov jump, etc.), we focus our attention solely on the chemical master equation (CME) dynamics, the setting where FFS was first developed \cite{allen2005sampling}.
More specifically, we study discrete reaction-diffusion dynamics of \(L\) sites or voxels as they are often called in the RDME literature~\cite{isaacson2006incorporating,erban2007practical}.
The rationale for this focus is twofold.
From a practical perspective, the RDME is important as a popular method to model chemical and biological processes \cite{fange2010stochastic,wilkinson2018stochastic,smith2019spatial}. 
From a theoretical perspective, the RDME is interesting because it provides a setting where one can analyze the influence of stochasticity on the emergence of patterns \cite{craciun2006understanding,krause2021modern,kondo2010reaction}.
In many pattern-formation problems, it is possible to understand the underlying physics in terms of a mean-field partial differential equation that tracks spatiotemporal evolution of deterministically evolving fields~\cite{turing1952chemical, cross1993pattern}.
It is, however, well-appreciated that stochastic effects can quantitatively and even qualitatively impact kinetics~\cite{mcquarrie1967stochastic,lee2006non,mcadams1997stochastic,thattai2001intrinsic,vilar2002mechanisms,paulsson2004summing,lee2006non}.
It is therefore important that the methodology we discuss captures more than just the typical dynamics.
Accurately computing transition rates between metastable states requires quantitatively resolving even rare tails of the state space distribution since these rare fluctuations can be instrumental in triggering transitions~\cite{tian2006stochastic,allen2006simulating}.

Historically, similar schemes for the CME ensemble evolution have been hampered by the curse of dimensionality. 
As a result, exact solutions are rare \cite{heuett2006grand} and numerical solutions often resort to sampling methods built upon the kinetic Monte-Carlo stochastic simulation algorithm (SSA) \cite{gillespie2013perspective,cao2006efficient}, also known as the Gillespie algorithm. 
Another approach is to evolve a subset of the state space through the finite state projection (FSP) method~\cite{munsky2006finite}. 
FSP has previously been combined with TNs to approximately describe the CME \cite{kazeev2014direct,vo2017adaptive}, resembling our use of TNs to tame the many-body problem.
Unlike those projective methods, however, the approach we describe uses the time-dependent variational principle (TDVP) to evolve the dynamics with probability conservation at all times, akin to prior work with diffusion and no reactions~\cite{helms2019dynamical,strand2022using,strand2022computing}.
Since we estimate rare events by measuring the distribution's small probability fluxes, conservation of probability and high-fidelity dynamical evolution are materially important.

While the formalism we describe applies quite generally to different well-mixed and reaction-diffusion models, we center our paper around a single model problem: rate calculations for transitions between basins of the Schl\"ogl model~\cite{schlogl1972chemical} in the bistable regime, both with and without diffusion~\cite{kim2017stochastic}. 
In Sec.~\ref{sec:wellmixed}, we first illustrate the rare event problem and extract rate constants from a propagator in the well-mixed regime.
For more complicated systems (either systems with more chemical species or more voxels), the curse of dimensionality precludes the exact calculation of the propagator, but the remainder of the paper shows how to employ TNs to follow the same conceptual path.
In Sec.~\ref{sec:dp}, we review the Doi-Peliti framework for writing reaction-diffusion dynamics in a second-quantized field theoretic form.
Sec.~\ref{sec:tn} then reviews the TN matrix product state (MPS) ansatz, a controllable approximation to the ensemble that can be evolved via TDVP.
Those methods are applied to the many-body rate calculation in Sec.~\ref{sec:mbrates}.
In Sec.~\ref{sec:Results}, we show results for the RDME with diffusion between \(L\) well-mixed Schl\"ogl voxels, demonstrating that the methodology can extend numerical transition rate calculations to regimes where the SSA becomes impractical.
The approach effectively evolves the probability of occupying each microstate in an unfathomably large many-body state space, yet the calculations scale sub-exponentially with system size and do not assume detail-balanced equilibrium.
Sec.~\ref{sec:Deploying} discusses how the TDVP calculations yield not only a rate, but also a mechanism.
Finally, in Sec.~\ref{sec:Discussion}, we discuss potential applications and challenges that should motivate future work.

\section{Ensemble rate calculations}\label{sect:EstRate}
\label{sec:wellmixed}

Consider an archetypal two-state rate process between reactants and products $\ce{R <=> P}$, initiated entirely in the reactant state.
After some reaction timescale, the concentration of products grows before eventually leveling off at its steady-state value.
The reaction rate can simply be defined and measured in terms of how quickly the product concentration rises.
An alternative formulation is to study the fate of an individual reactant on its path to transitioning into a product.
The TPS and FFS path sampling approaches extract the reaction rate by focusing on the statistical properties of these individual trajectories.
Algorithms exist to propagate those single trajectories, algorithms that can be practically implemented even when the system grows large and complex.
By contrast, propagating the entire ensemble of trajectories has typically been limited to simple low-dimensional problems.
In this section, we illustrate an ensemble-evolution rate calculation for a simple case in order to develop notation and lay the groundwork for the many-body problem.
The remainder of the paper will show that the methodology can be extended beyond such simple problems by employing TNs.

\subsection{Well-mixed Schl\"ogl Model}
\label{sec:WMSchlogl}
\begin{figure}[t!]
\centering
\includegraphics[width=.45\textwidth]{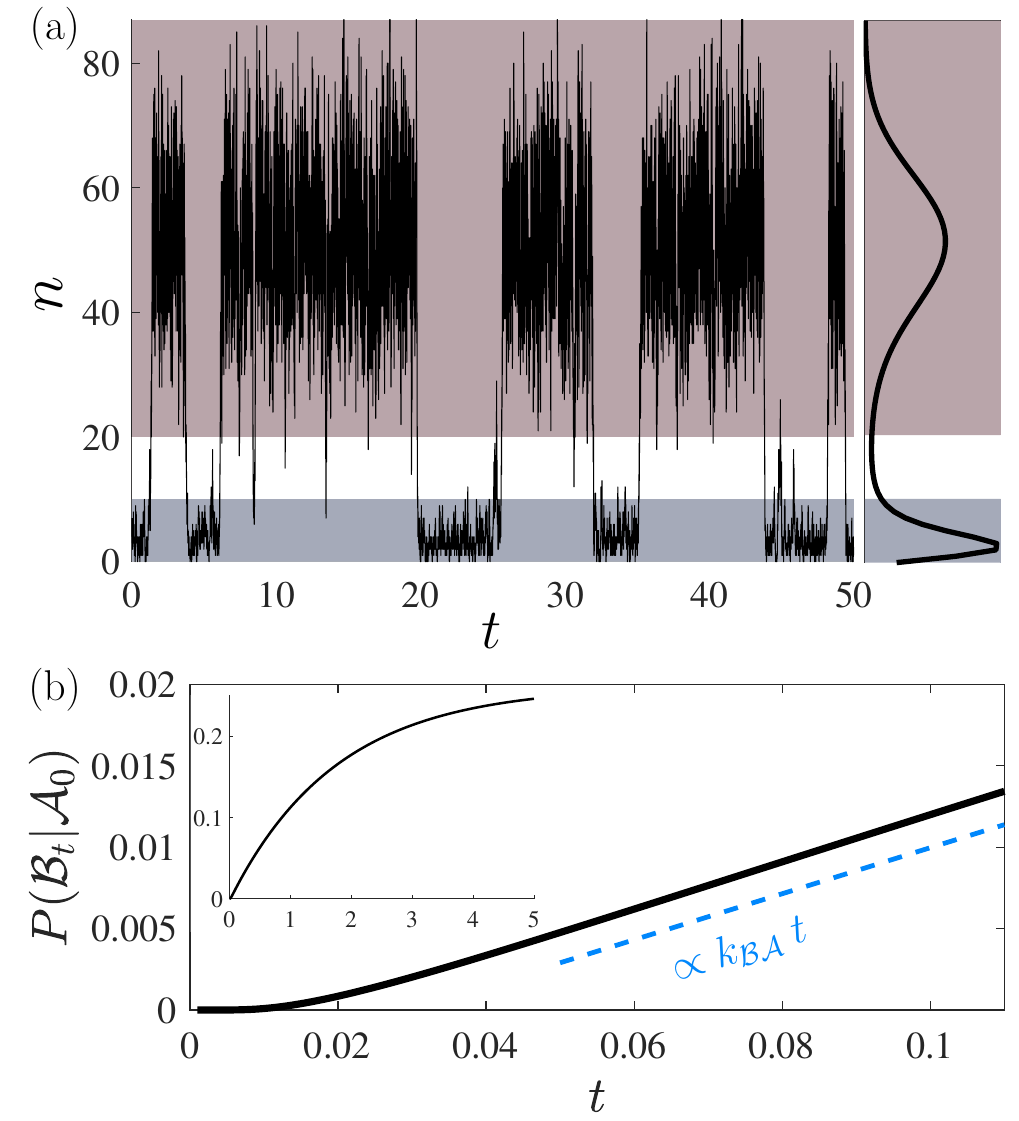}
\caption{\label{fig:ER}(a) A typical trajectory in the bistable regime ($\tilde{c}_1 = 36.38, \tilde{c}_2 = 0.040, \tilde c_3 = 2200,$ and $\tilde c_4 = 37.6$) of the well-mixed Schl\"{o}gl model of Sec.~\ref{sec:WMSchlogl}.
  The system stochastically switches between the two metastable basins, where basin $\mathcal{A}$ is the top colored region and $\mathcal{B}$ is the bottom colored region.
  (b) Starting with a distribution localized in basin $\mA$, the evolution has three time scales.
  Initially, no trajectories can reach $\mB$.
  After the so-called molecular timescale $\tau_{\rm mol}$, the fastest trajectories begin to reach $\mB$.
  Subsequently, probability leaches into region $\mB$ at a constant rate with $P(\mB_t|\mA_0) \propto k_{\mB\mA}$.
  The inset shows that growth on a much longer timescale, revealing the third scale when the system eventual approaches the steady state.
}
\end{figure}

Consider the well-mixed Schl\"ogl model, a quintessential example of bistability in chemical reaction networks (CRNs)~\cite{vellela2009stochastic,nicholson2018entrance}.
The model tends to be studied under chemostatted conditions that fix two species' concentrations $A$ and $B$, while allowing species X's concentration to evolve according to the four reactions:
\begin{align}
\ce{$2$X + A &<=>[\tilde{c}_1][\tilde{c}_2] $3$X}\label{rxn:S1} \\ 
\ce{B &<=>[\tilde{c}_3][\tilde{c}_4] X}
\label{rxn:S2}
\end{align}
where $\tilde{c}_1,\tilde{c}_2,\tilde{c}_3$ and $\tilde{c}_4$ are the stochastic rate coefficients.
Our starting point is microscopic such that the rate \(\tilde{c}_1\) should be interpreted as the reciprocal of the waiting time for a particular set of $2$ X molecules to react and form 3 X molecules when mediated by a fixed concentration of A molecules.
In the large-system, well-mixed limit of chemical kinetics, it is common to also define macroscopic, or kinetic rate constants \(k_r = \tilde{c}_r (N_{\rm A} V)^m\) for each reaction \(r\) with \(m\) reactants.
Here, Avogadro's number \(N_{\rm A}\) serves to convert between microscopic measures (counting molecules) and macroscopic kinetic measures (concentrations in molarity) for a volume \(V\).
In this paper, we focus on the microscopic kinetic picture, viewing the $\tilde{c}_r$ as the fundamental parameters, but these parameters can be converted into $k_r$ for linear reactions or in the thermodynamic limit~\cite{higham2008modeling}.
In that thermodynamic limit, the stochastic kinetics often tends to the solutions of the deterministic reaction rate equations.
For the Schl\"{o}gl model, analysis of that deterministic limit reveals that the sign of the discriminant
\begin{align}
\Delta &= 4k_1^3 k_3A^3B - k_1^2k_4^2A^2 + 4k_2k_4^3 - 18k_2k_1k_4k_3AB \nonumber \\
&\ \ \ + 27k_2^2k_3^2B^2
\end{align}
determines the number of stable steady-state solutions~\cite{vellela2009stochastic}.
While holding the rate coefficients fixed, the sign of $\Delta$ changes as a function of the concentrations of A and B, allowing one to pass from $\Delta > 0$ with one steady-state solution, through $\Delta = 0$ with a diverging variance of the number of X molecules~\cite{nicolis1977stochastic}, and on to $\Delta < 0$ with one unstable and two stable solutions.
In the \(\Delta <0\) regime, a finite (stochastic) system sees the regions around the two stable fixed points become metastable states.
The number of X molecules stochastically switches between the two metastable values.
A representative trajectory, shown in Fig.~\ref{fig:ER}, oscillates between a state with \(n \approx 5\) X molecules and another with \(n \approx 50\).
We first set out to compute the rate of bistable switching between those two metastable states in the well-mixed regime, making use of the fact that the dynamics is one-dimensional (1D)\textemdash \(n\) executes a random walk between 0 and some large maximum number \(M\), which we impose for the sake of calculation.

Rather than track \(n(t)\) for a single trajectory as in Fig.~\ref{fig:ER}, the ensemble approach tracks evolution of the distribution \(p_t(n)\), measuring the probability that an individual trajectory would have \(n\) molecules at time \(t\).
Foreshadowing the analogies with quantum-mechanical methodologies, we choose to write that distribution as a ket: \(\ket{p_t} = \left[p_t(0), p_t(1) \hdots, p_t(M)\right]^T\), where \(T\) indicates the transpose to give a column vector.
The distribution evolves according to the master equation
\begin{equation}
\frac{d\ket{p_t}}{dt} = H\ket{p_t}.
\label{eq:ClSch}
\end{equation}
Here, \(H\) is a rate matrix whose off-diagonal elements \(H_{n'n}\) are the rates of transitioning from \(n\) X molecules to \(n'\) and whose diagonal elements \(H_{nn}\) are minus the rate of escaping from the state with \(n\) X molecules.
The rate matrix conserves probability, a fact compactly expressed as $\bra{\mathbf{1}}H = 0\bra{\mathbf{1}}$, with $\bra{\mathbf{1}}$ denoting a vector of ones.
The distribution \(\ket{p_t}\) can always be formally evolved in terms of a propagator \(e^{Ht}\) as 
\begin{equation}
\ket{p_t} = e^{Ht}\ket{p_0}.
\label{eq:Propagator}
\end{equation}
Due to the analogy with quantum-mechanical time evolution, we equivalently call \(H\) an effective Hamiltonian.
Assuming $H$ is irreducible, the long-time limit for this distribution tends to the unique steady-state $\ket{\boldsymbol{\pi}} = \lim_{t\to \infty} e^{Ht}\ket{p_0}$. 

\(H\) can be decomposed into contributions from the two different reversible mechanisms of Eqs.~\eqref{rxn:S1} and~\eqref{rxn:S2}, labeled \(H^{A}\) and \(H^{B}\), respectively, based on whether the transitions are mediated by the chemostatted species A or B.
Then, the Schl\"ogl model at the level of the master equations corresponds to tridiagonal rate matrices with elements
\begin{align}
  \label{eq:tridiagonal}
H^A_{n'n} = &\frac{\tilde{c}_1 n (n-1)}{2}\delta_{n'-1,n} + \frac{\tilde{c}_2 n (n-1)(n-2)}{6}\delta_{n'+1,n} - \nonumber \\
&\frac{n (n-1)}{2}\left(\tilde{c}_1 + \frac{\tilde{c}_2(n-2)}{3}\right)\delta_{n',n}
\end{align}
and
\begin{equation}
H^B_{n',n} = \tilde{c}_3\,\delta_{n'-1,n} + \tilde{c}_4n \delta_{n'+1,n} - \left(\tilde{c}_3 + \tilde{c}_4n \right)\delta_{n',n},
\end{equation}
where $H = H^A + H^B$ and $\delta_{i,j}$ is the Kronecker delta.
Given the specific tridiagonal \(H\) and the restriction \(0 \leq n \leq M\), both \(H\) and the propagator \(e^{Ht}\) are \((M+1) \times (M+1)\) matrices, amenable to evolving the entire distribution, effectively averaging over all the realizations.

To extract a rate from the evolution of distribution, we must initialize the distribution in one of the two metastable states and measure the rate of relaxation.
In analogy with free energy basins, we will call the two metastable regions basins \(\mathcal{A}\) and \(\mathcal{B}\) (red and blue regions of Fig.~\ref{fig:ER} respectively) with \(\mathcal{A} \cap \mathcal{B} = \emptyset\).
We write \(\proj_{\omega}\) for the non-negative idempotent ($\proj_{\omega}^2 = \proj_\omega$) operator that projects onto the density lying in a region \(\omega\), where \(\omega \in \{\mathcal{A},\mathcal{B}\}\).
If \(\ket{p_t}\) has support over a state space \(\Omega\), then the projection \(\proj_\omega \ket{p_t}\) is an unnormalized distribution with support restricted to \(\omega\).
The total probability in \(\omega\) at time \(t\) can then be expressed as \(\bra{\mathbf{1}}\proj_\omega\ket{p_t}\).

\subsection{Rate Constants}
\label{sec:rateconstants}
In general, one can compute the probability to pass from any subset $\mA \subset \Omega$ into an arbitrary non-overlapping subset $\mB$.
For the current example, $\mathcal{A}$ and $\mathcal{B}$ are basins of attraction for the Schl\"{o}gl model (see Fig.~\ref{fig:ER}a), and we would like to measure the rate of going from $\mathcal{A} \rightarrow \mathcal{B}$.
Rate constants can be defined either for the unidirectional flux from $\mathcal{A} \rightarrow \mathcal{B}$, the quantity we focus on in this work, or for the timescale of the relaxation to the steady-state, which also involves the $\mathcal{B} \rightarrow \mathcal{A}$ transitions~\cite{roux2022transition}.
Additionally, the precise definition of the rate depends on whether one seeks trajectories that move from the boundary of one basin to another or from the interior of one basin to another.
For instance, studying the time to transition from the boundary of \(\mathcal{A}\) to the boundary of \(\mathcal{B}\) yields the Transition Path Theory (TPT) rate associated with the committor function~\cite{e2010transition}.
This unidirectional rate $k_{\mB \mA}$ is related to the probability of having reached $\mB$ from $\mA$ in time $t$ without returning to $\mA$.
If we focus on trajectories short enough that those reaching $\mB$ will stay (no recrossings) then $k_{\mB \mA}$ can be expressed in terms of the probability $P(\mA_0)$ to start in $\mA$ at time zero and the conditional probability $P(\mB_t | \mA_0)$ to subsequently be found in $\mB$ at time $t$:
\begin{equation}
k_{\mathcal{B} \mathcal{A}} \equiv \frac{d}{dt} P(\mB_t | \mA_0)P(\mA_0)\biggr\rvert_{t >\tau_{\rm mol}}.
    \label{eq:rate_a}
\end{equation}
The derivative is evaluated at some timescale exceeding $\tau_{\rm mol}$, a molecular time scale of the system that is long enough for the first trajectories to reach \(\mB\) from \(\mA\) but much shorter than the typical transit time $k_{\mathcal{B}\mathcal{A}}^{-1}$ (see Fig.~\ref{fig:ER}b).

We compute the rate of Eq.~\eqref{eq:rate_a} by initializing in a distribution confined to $\mA$ then propagating for time $t$ and measuring the probability in $\mB$.
A suitable initial distribution is $\proj_\mA\ket{\boldsymbol{\pi}}/\bra{\mathbf{1}}\proj_{\mA}\ket{\boldsymbol{\pi}}$, a normalized distribution constructed by projecting the steady-state distribution $\ket{\boldsymbol{\pi}}$ onto $\mA$.
With this contribution, $P(\mA_0)=1$, and Eq.~\eqref{eq:rate_a} becomes
\begin{equation}
  k_{\mathcal{B} \mathcal{A}} =  \frac{d}{dt} \frac{\bra{\mathbf{1}}\proj_{\mB}\,e^{Ht}\, \proj_{\mA} \ket{\boldsymbol{\pi}}}{\bra{\mathbf{1}}\proj_\mA\ket{\boldsymbol{\pi}}}\biggr\rvert_{t >\tau_{\rm mol}}.
    \label{eq:rate}
\end{equation}
To the extent that there is a separation of timescales between exploring a basin and transiting between basins, the transition event is rare and no transitions can occur faster than the molecular time scale \(\tau_{\rm mol}\).
Beyond that time, the number of transitions accumulates over time as \(P(\mathcal{B}_t|\mathcal{A}_{0}) \propto k_{\mathcal{B}\mathcal{A}}  t\).
At times long compared to \(k_{\mB \mA}^{-1}\), \(P(\mathcal{B}_t|\mathcal{A}_{0})\) must eventually stop growing linearly to level off at the stationary value \(\bra{\mathbf{1}}\proj_\mB \ket{\boldsymbol{\pi}}\), but our interest is in extracting \(k_{\mB \mA}\) from the growth at much shorter times.
A typical realization of the process will not require observation much beyond \(\tau_{\rm mol}\) units of time, to ascertain the rate, meaning our ensemble evolutions must be propagated for roughly $\tau_{\rm mol}$.
Notice from the inset of Fig.~\ref{fig:ER}(b) that at long times the recrossing events obscure the rate of growth, ultimately yielding the plateau.
As discussed in Sec.~\ref{sec:mbrates} and App.~\ref{app:modifieddynamics}, for rare rates and long times, the rate calculations can be improved by removing these recrossing events via a modified dynamics in which region $\mB$ is an absorbing state, forbidding dynamics out of that region.

Our ability to compute \(k_{\mB \mA}\) from the ensemble evolution traces back to the fact that it is practical to numerically calculate \(e^{Ht}\) for the well-mixed model with a single species and a modest \(M\).
While Eq.~\ref{eq:rate} generalizes naturally for harder problems (those with more than one species or with diffusion between multiple well-mixed voxels), the state space for \(\ket{p_t}\) grows exponentially with additional stochastic degrees of freedom.
We next show that the Doi-Peliti framework re-expresses the rate matrix in a second-quantized form that allows us to generalize beyond the case of a single well-mixed species.
 
\begin{figure*}[!ht]
\centering
\includegraphics[width=0.95\textwidth]{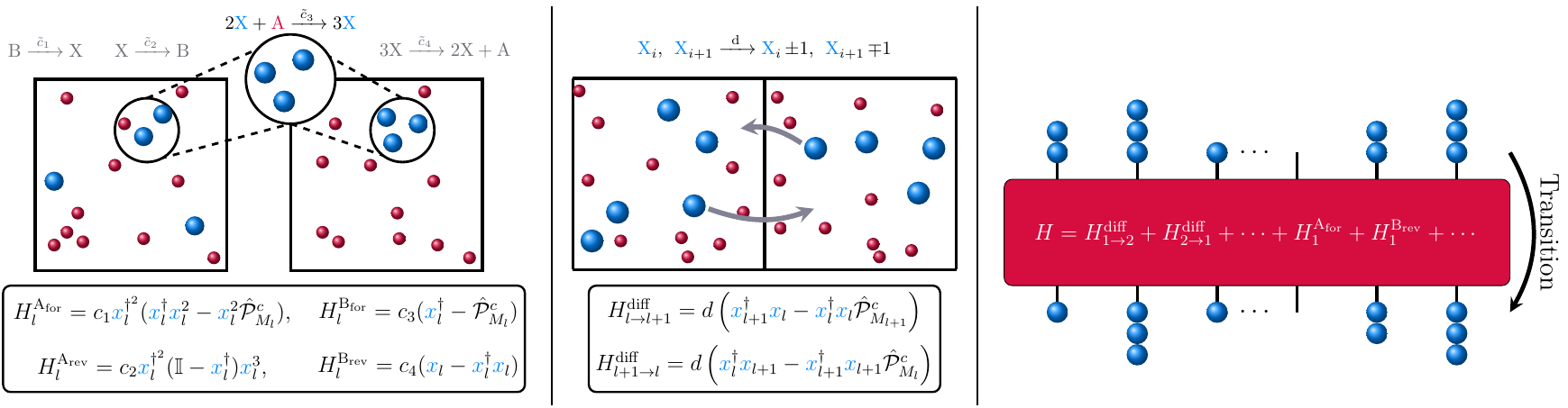}
\caption{\label{fig:one} The reaction-diffusion master equation (RDME) consists of voxels in which well-mixed chemical reactions occur and between which species can diffuse.
  This work centers around a one-dimensional chain of $L$ such voxels, each with a fluctuating number of molecules of species X in a Schl\"{o}gl chemical reaction.
  The Doi-Peliti (DP) framework turns the reaction-diffusion evolution into a second-quantized effective Hamiltonian that governs the evolution of the distribution over classical reaction-diffusion microstates.
  That effective Hamiltonian is represented by a high-dimensional tensor, with each vertical leg taking one of $(M+1)$ possible values corresponding to the number of molecules of species X which could occupy each site.
  The DP form of the effective Hamiltonian lends itself to a compact representation of the effective Hamiltonian tensor of rates for transitioning from one microstate to another.}
\end{figure*}

\section{Doi-Peliti}

\label{sec:dp}
The Doi-Peliti (DP) formalism is a classical version of the second quantization methods from quantum field theory \cite{ohkubo2013algebraic,tauber2014critical}.
Originally introduced by Doi \cite{doi1976stochastic} as a critique to the assumptions of the Smoluchowski equation, it was later developed by Peliti \cite{peliti1985path} and others \cite{rey1999asymptotic} as a path integral formalism for solving chemical reaction networks \cite{reyes2023general,vastola2021solving,yang2017analytical,harsh2023accurate}. The DP framework has also found success in other areas such as studying kinetically constrained lattice models in the context of glassy dynamics~\cite{garrahan2007dynamical,garrahan2009first}, cellular signaling \cite{lan2006variational}, predator prey models \cite{de2022dynamical} and active matter \cite{bothe2021doi}.
The DP representation builds operators which encode the dynamics onto the high-dimensional space that accompanies many-body problems.
Crucially, writing the chemical reactions in an operator form guarantees it can be easily converted into a matrix product operator (MPO)~\cite{schollwock2011density,causer2022finite}, as we will do in Sec.~\ref{sec:tn}.\\

To start, we go beyond Sec.~\ref{sec:wellmixed} by considering well-mixed reactions with multiple dynamic species  as well as heterogeneous systems where molecules can diffuse between neighboring voxels.
Let $X_1, X_2,\ldots X_L$ represent each chemical species in each site.
We could be talking about a single species \(X\) with different values in \(L\) different voxels, a single well-mixed voxel with \(L\) distinct dynamic species, or a mixture with multiple species and multiple voxels.
The number of molecules in a given voxel can change due to a chemical reaction or a diffusion event, with both events being cast as,
\begin{equation}
\ce{ $\sum_j \eta_j^r$ X$_j$ ->[\tilde{c}_r]  $\sum_j \mu_j^r$ X$_j$}
\label{eq:rxn}
\end{equation}
where $\tilde{c}_r$ are again the rate coefficients for reaction $r$.
$\eta_j^r$ and $\mu_j^r$ reflect the stochiometry of the reactions and give the stochiometric vector $\nu^r = \mu^r - \eta^r$ which evolves the system from state $\bfn \in \Omega$ to $\bfn' = \bfn + \nu^r$.
In the previous section, these reactions induced changes between the \(M+1\) different microstates, specified by the scalar \(n\).
Now, the microstates $\bfn$ are specified by \(\mathbf{n} = [n_1, n_2, \hdots, n_L]\), counting the occupation of each species within each voxel. 

For a dilute well-mixed solution, where well-mixed means molecules locations within a voxel are unresolved, the change in probability of microstate $\bfn$ follows the gain-loss equation \cite{isaacson2006incorporating},
\begin{equation}
\frac{dp_t(\bfn)}{dt} = \sum_{r}^N\left[ \alpha_{\bfn - \nu^r}^r p_t({\bfn - \nu^r}) - \alpha_{\bfn}^r p_t({\bfn})\right],
\label{eq:chgPr}
\end{equation}
where the sum is taken over \(N\) different reactions of the form Eq.~\eqref{eq:rxn}, each with their own stochiometric vector \(\nu^r\).
The transition rate between any two states is the sum of contributions from each reaction \(H_{\bfn'\bfn} = \sum_r \alpha^r_{\bfn}\).
Each propensity $\alpha^r$ is a product of a stochastic rate coefficient $\tilde{c}_r$ and the combinatorial number of ways to select either $\eta_j$ or $\mu^r$ molecules from reaction $r$, depending on the direction of the reaction.
For a ``forward" reaction the propensity is, $\alpha_r(\bfn) = \tilde{c}_r\prod_{n_j\in r}{n_j\choose\eta_j}$.

Encoding the reactions into a matrix form allows us to evolve the elements of the joint distribution $p_t(\bfn)$ analogously to the previous section.
While it is appealing to try and directly solve these more complicated dynamics as we did before, it is impractical.
The curse of dimensionality demands a less onerous way to work with both \(H\) and \(p_t(\bfn)\).
Since particles of the same species in the same voxel are indistinguishable, it is beneficial to work in an occupation number representation with
\begin{equation}
    \ket{p_t} = \sum_{\bfn} p_t(\bfn) \ket{\bfn}.
\end{equation}
The basis vectors
\begin{align}
    \nonumber \ket{\bfn} &= \ket{n_1, n_2, \hdots n_L}\\
    &= \ket{n_1} \otimes \ket{n_2} \otimes \hdots \otimes \ket{n_L}
\end{align}
corresponds to a possible microstate, and the set of basis vectors forms a tensor product space called a Fock space $\mathcal{F}^L$ with $\ket{\bfn} \in \mathcal{F}^L$ and \(\bra{\mathbf{m}}\ket{\bfn} = \delta_{\mathbf{m}\bfn}\)~\cite{del2022probabilistic}.
Without loss of generality, we will discuss the case of a single chemical species and \(L\) physical voxels, in which case \(n_l\) has the interpretation of the number of molecules in voxel \(l\).
The second-quantization procedure rewrites the rate matrix as an operator \(\hat{H}\) in terms of creation and annihilation operators, \(x_l^\dagger\) and \(x_l\), that respectively raise and lower the number of X molecules in voxel \(l\).
That is to say,
\begin{align}
x_l^\dagger\ket{n_l} &= \ket{n_l+1},\nonumber \\
x_l\ket{n_l} &= n_l\ket{n_l-1},\nonumber \\
x_l\ket{0} &= 0.
\label{eq:raiselower}
\end{align}
where $\ket{0}$ is the vacuum state, i.e., no molecules.
By $x_l$ or $x_l^\dagger$ we are implying that the operator only acts on the $l^{\rm th}$ voxel:
\begin{equation}
  x_l = \mathbb{I}_1 \otimes \mathbb{I}_2\otimes\ldots\otimes \mathbb{I}_{l-1}\otimes  x \otimes \mathbb{I}_{l+1} \otimes \ldots \otimes \mathbb{I}_L
\end{equation}
Because they act on the single-site space, the general one-site operators $x$ and $x^\dagger$ can be written in a matrix form as
\begin{equation}
x^\dagger = \begin{bmatrix} 
0 && 0 && 0 && \ldots \\
1 && 0 && 0 && \ldots \\
0 && 1 && 0 && \\
\vdots && && \vdots && \ddots
\end{bmatrix} \text{ and }
\qquad
x = \begin{bmatrix} 
0 && 1 && 0 && \ldots \\
0 && 0 && 2 && \ldots \\
0 && 0 && 0 && \\
\vdots && && \vdots && \ddots
\end{bmatrix}.
\label{eq:xmatrices}
\end{equation}
That these matrices yield Eq.~\eqref{eq:raiselower} can be readily confirmed.
More abstractly, the existence of such creation and annihilation operators in this classical setting originates from the classical commutation relation \([x_i, x_j^\dagger] = \delta_{ij}\) and the existence of a positive semi-definite number operator $\hat{n}_l = x_l^\dagger x_l$, so named because \(\hat{n}_l\ket{n_l} = n_l \ket{n_l}\).
Together, the set $\{\Omega,\mathcal{F}^L,x,x^\dagger\}$ are also sometimes called an interacting Fock space \cite{ohkubo2013algebraic}.
As in the construction of the number operator, we will be able to decompose contributions to \(\hat{H}\) in terms of the \(x^\dagger\) and \(x\)'s.

In practice, the matrices of Eq.~\eqref{eq:xmatrices} will not be infinite in size but will rather be \((M+1) \times (M+1)\) matrices due to our enforcement of a maximum voxel occupancy of $M$.
At the operator level, this maximum voxel occupancy supplements Eqs.~\eqref{eq:raiselower} with an additional requirement that
\begin{equation}
  x_l^\dagger \ket{M_l} = 0,
  \label{eq:cap}
\end{equation}
a requirement that is met by the finite truncation of Eqs.~\eqref{eq:xmatrices}.
Those truncated operators obey a commutation relation which is similar to that of their infinite counterpart, but the usual commutation relation is broken at the maximal occupancy state.
That is to say
\begin{equation}
  (x_l x_l^\dagger - x_l^\dagger x_l) \ket{M_l} = - M_l \ket{M_l} \neq \ket{M_l}.
\end{equation}
The commutation relation for the truncated operators is therefore
\begin{equation}
  [x_i, x_j^\dagger] = \delta_{ij}\bigg(1 - \left(M+1\right)\ket{M_i}\bra{M_i}\bigg),
  \label{eq:realcomm}
\end{equation}
an expression involving the projector operator onto site $i$'s maximal occupancy state, $\hat{\mathcal{P}}_{M_i} = \ket{M_i}\bra{M_i}$.
As discussed in depth in App.~\ref{app:DPEnc}, this violation of the usual commutator relation must be accounted for when constructing an effective Hamiltonian that conserves probability.

\begin{figure*}[!ht]
\centering
\includegraphics[width=.9\textwidth]{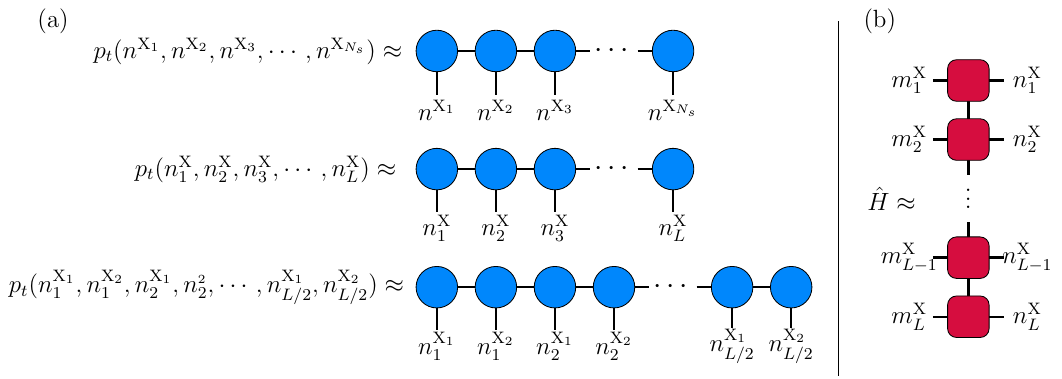}
\caption{\label{fig:mps} The ensemble evolution of the many-body state requires that one solve for the probability of every microstate at all times.
  That high-dimensional joint probability distribution can be well approximated by a matrix product state (MPS).
  An MPS approximation for the joint distribution can be advantageous whether there are $N_s$ species in a well-mixed reaction [(a) top], one dynamical (unchemostatted) species with diffusion [(a) middle], or multiple species and diffusion [(a) bottom].
  In all three cases, each circle represents a tensor, either a rank-$2$ at the first and last site or rank-$3$ in the body.
  Horizontal lines represent the bond-indices connecting tensors and vertical lines are physical indices.
  For an RDME, the physical indices represent the possible number of molecules at a given tensor.
  If two lines are connected, then the index is summed over.
  (b) An effective Hamiltonian $H$ can be likewise decomposed into lower-rank tensors to form a matrix product operator (MPO).
  Now there are two physical indices per tensor, representing transitions from state $n_l^X$ to $m_l^X$.
  Vertical connected lines are the bond indexes which are contracted over.}
\end{figure*}

One benefit of defining operators in terms of creation and annihilation operators is that expectation values for observables can be expressed easily in terms of inner products with operators.
Consider, for example, the observable that counts all particles in all voxels, \(\hat n = \sum_l \hat n_l\).
The operator \(\hat{n}\) acts on a many-body state \(\ket{\bfn}\), yet each \(\hat{n_l}\) is a single-body operator acting on voxel $l$ and leaving all other voxels unchanged.
It is therefore understood that when local operators such as $\hat n_l$ act on $\ket{\bfn}$, it is written as a shorthand for
\begin{equation}
\hat n_l = \mathbb{I} \otimes \hdots \otimes x^\dagger_lx_l \otimes \hdots \otimes \mathbb{I}.
\end{equation}
In terms of that shorthand, $\hat n_l \ket{\bfn} = n_l \ket{\bfn}$ counts the number of particles in voxel $l$ for the many-body state $\ket{\bfn}$.
The time-dependent expected number of particles in the whole system is thus \(\langle n(t) \rangle = \sum_l\bra{\mathbf{1}}\hat n_l \ket{p_t}\), where the vector of ones \(\bra{\mathbf{1}}\) serves to sum over all microstates.
The time-dependence of a distribution can also be extracted from the action of a second-quantized \(\hat{H}\) on \(\ket{p_t}\), and this operation can be practically computed when \(\hat{H}\) is expressed in terms of the single-site operators.

Mirroring the discussion in Sec.~\ref{sec:wellmixed}, the generator $\hat H_l$ for the Schl\"ogl model reactions within the  $l^{\rm th}$ voxel is a sum of contributions from the elementary reactions $\hat H_l = \sum_r \hat H^r$.
Fig.~\ref{fig:one}a shows each reaction, expressed as a second-quantized contribution to the effective Hamiltonian, as derived in App.~\ref{app:DPEnc}.
Probability conservation requires that this effective Hamiltonian include a term with a projector $\proj^c_{M_l} \equiv \mathbb{I} - \hat{\mathcal{P}}_{M_l}$ that prevents flows of probability into forbidden microstates with occupancy greater than $M$.
This projector is complementary to the one appearing in Eq.~\eqref{eq:realcomm} in that it projects onto states which are \emph{not} maximally occupied.
Combining together the forward and reverse reactions mediated by species A, we show in App.~\ref{app:DPEnc} that the reversible reaction Eq.~\eqref{rxn:S1} contributes 
\begin{equation}
  \hat{H}^A_l = c_1\left[ x_l^{\dagger 3}x_l^2 - x_l^{\dagger 2}x_l^2 \proj^c_{M_l}\right] + c_2\left[ x_l^{\dagger 2}x_l^3 - x_l^{\dagger 3}x_l^3 \right]
  \label{eq:HA}
\end{equation}
to the effective Hamiltonian.
For convenience we have absorbed the combinatorial terms into that rate coefficient, defining $c_r \equiv \tilde{c}_r / \gamma!$, where $\gamma$ is either $\eta^r$ or $\mu^r$ for forward and reversed reactions, respectively.
Likewise, the contribution
\begin{equation}
  \hat{H}_l^B = c_3 \left(x_l^\dagger - \proj^c_{M_l}\right) + c_4 \left(x_l - x_l^\dagger x_l\right).
  \label{eq:HB}
\end{equation}
comes from Eq.~\eqref{rxn:S2}.
In the $M \rightarrow \infty$ limit, $\proj^c_{M_l} \to \mathbb{I}$ to give the more familiar expressions written in Fig.~\ref{fig:one}a.

The Hamiltonian operator for \(L\) independent, well-mixed Schl\"ogl voxels is thus \(\hat{H}^{\rm rxn} = \sum_{l}^L \hat{H}^A_l + \hat H^B_l\).
To add diffusion, consider the well-mixed voxels arranged in a 1D lattice as in Fig.~\ref{fig:one}b.
Each chemical species is able to move between neighboring voxels as regulated by a diffusion operator
\begin{align}
  \hat{H}^{\rm diff} = &\sum_{l =1}^{L-1} d\left(x^\dagger_{l+1}x_l - x^\dagger_lx_l\proj^c_{M_{l+1}}\right) \nonumber \\
  &\ \ \ \ \ + d\left(x^\dagger_l x_{l+1} - x^\dagger_{l+1}x_{l+1}\proj^c_{M_l}\right).
  \label{eq:Hdiff}
\end{align}
with hopping rate \(d = D/h^2\), consistent with macroscopic diffusion constant $D$ for voxel width $h$  \cite{smith2019spatial}.
Combining reaction with diffusion gives
\begin{equation}
  \hat{H} = \hat{H}^{\rm rxn} + \hat{H}^{\rm diff},
\end{equation}
a straightforward linear superposition at the level of the effective Hamiltonian.
We emphasize that the linear superposition of diffusion $\hat{H}^{\rm diff}$ with reactions $\hat{H}_l^A$ and $\hat{H}_l^B$ does not imply a restriction to linear kinetics.
The Schl\"ogl model kinetics is nonlinear in that it involves elementary reactions which are not first order, and the diffusive coupling between neighbors can induce nontrivial interactions between voxels.

\section{Tensor Networks}\label{sec:tn}

We have now seen how to transform the matrix-form rate operator $H$ of Sec.~\ref{sect:EstRate} into a Doi-Peliti form $\hat{H}$, expressed in terms of local creation and annihilation operators.
That change of representation is not merely cosmetic.
As the number of chemical species or number of lattice sites grows, the size of $H$ grows exponentially.
For $N$ dynamical (unchemostatted) species and $L$ lattice sites, the state space has $(M+1)^{NL}$ states.
To put this scaling in perspective, the largest numerical result that follows uses $N=1$, $M=85$ and $L=8$, meaning the system contains $2.99\times 10^{15}$ microstates.
Calculations on a matrix $H$, even a very sparse matrix $H$, are untenable.
By contrast, it is practical to construct a TN decomposition of the second quantized form of $\hat H$ in terms of an MPO, as shown in Fig.~\ref{fig:mps}b, and demonstrated explicitly for the Schl\"ogl model in App.~\ref{app:MPO}.
Given the second quantized form, the MPO can be readily constructed in a practical form through SVD decomposition using automated packages such as AutoMPO in ITensor~\cite{itensor}.

Much like the operator \(\hat{H}\) can be compressed as an MPO, the states that operators act upon can also be expressed more compactly in terms of TNs.
Consider \(p_t(\bfn) = p_t(n_1, n_2, \hdots, n_L)\), the probability of microstate \(\bfn\) at time \(t\).
This $p_t(\bfn)$ can be thought of as a rank \(L\) tensor.
By specifying the occupancy of each of the \(L\) sites, $p_t(\bfn)$ outputs a number\textemdash a probability\textemdash associated to that microstate.
By leveraging a matrix product ansatz, $p_t(\bfn)$ can be approximated as a partial contraction over $L$ different low-rank tensors (see Fig.~\ref{fig:mps}a).
The low rank tensors of an MPS~\cite{orus2014practical} are organized in a 1D chain of site dependent tensors \(Q_{s_1}^{n_1}, Q_{s_1 s_2}^{n_2}, \hdots, Q_{s_{L-2} s_{L-1}}^{n_{L-1}},\) and \(Q_{s_{L-1}}^{n_L}\) whose upper index \(n_{l}\) specifies the occupancy of site \(l\) and whose lower indices are so-called bond indices, nonphysical indices which are contracted over to generate correlations between nearby sites.
For compactness of notation, we suppress labels $Q^{(1)}, Q^{(2)},$ etc. that would distinguish the site dependent tensors, implicitly carrying that site dependence through the labeling scheme for the physical indices.
For suitably chosen \(Q\) tensors, we therefore approximate
\begin{align}
  \ket{p_t} &= \sum_{n_1, n_2, \hdots n_L} p_t(n_1,n_2,\ldots,n_L) \ket{n_1, n_2, \hdots n_L} \nonumber\\
  &\approx \hspace{-3mm} \sum_{\substack{n_1,n_2, \hdots,n_L \\ s_1, s_2, \hdots,s_{L-1}}}\hspace{-3mm} Q^{n_1}_{s_1}Q^{n_2}_{s_1s_2}\ldots Q^{n_L}_{s_{L-1}}\ket{n_1, n_2, \hdots n_L} \nonumber\\
&\equiv \ket{q_t},
\label{eq:Pr_MPS}
\end{align}
where $\ket{q_t}$ is the approximate distribution over microstates generated from the MPS.
Though not explicitly written, the \(Q\) tensors are time-dependent.

There is a diagrammatic representation of the tensors, shown in Fig.~\ref{fig:mps}, which aids in visualizing tensor operations.
The tensors \(Q\) are given by circles, one per site, and vertical lines feeding into these circles are the physical indices that pick out a particular slice of the tensor based on the occupation number of that site.
Circles also have horizontal lines emanating from them, representing the non-physical bond indices.
Connection of two circles by a line represents a shared index being contracted over.
Fig.~\ref{fig:mps} emphasizes that the TN decomposition is flexible enough to address varied CME's.
The MPS of Fig.~\ref{fig:mps}a middle, has been the focus.
Each site from 1 to \(L\) reflects the number of X molecules in that site, but Fig.~\ref{fig:mps}a (top) and (bottom) show that the MPS decomposition can equally well be constructed when multiple species X$_i$, $i = 1,2,\ldots,N_{\rm s}$  can occupy a single site or when $N_{\rm s}$ different species diffuse between $L/N_{\rm s}$ sites to yield an MPS with $L$ sites.
What is practically important is that \(Q\)'s be arranged such that the physical dynamics correlates the neighbors.
For example, for the RDME Schl\"{o}gl model, neighboring \(Q\)'s correspond to neighboring voxels in the 1D Schl\"{o}gl lattice.

An attractive feature of the MPS ansatz is that the size and accuracy of the MPS are controllable through the bond dimension of the auxiliary indices $s$.
If the dimension of those indices is allowed to grow exponentially from $s_1$ through $s_{L/2}$, the ansatz in Eq.~\eqref{eq:Pr_MPS} is exact, but even when the bond dimension of the indices $s$ are capped at a maximum dimension of $\kappa$, the ansatz can be a very good approximation.
Crucially, capping that maximum bond dimension makes it practical to work with $\ket{q_t}$ because the rank-$L$ tensor $\ket{p_t}$ has been replaced by the $L$ different $Q$ tensors, each with no more than $(M+1)\kappa^2$ elements.

Having approximated both $\ket{q_t}$ and $\hat{H}$ by an MPS and MPO, respectively, we can now revisit the otherwise intractable dynamics problem of Eq.~\eqref{eq:ClSch}, only now we wish to evolve $\ket{q_t}$ in lieu of $\ket{p_t}$.
 As $\ket{q_t}$ is an MPS state, we wish for the dynamics to be constrained for all times to the manifold $\mathcal{M}$ of possible MPS states.
  Merely replacing $\ket{p_t}$ by $\ket{q_t}$ in Eq.~\eqref{eq:ClSch}:
\begin{equation}
  \frac{d \ket{q_t}}{dt} = \hat{H} \ket{q_t}
  \label{eq:qeom}
\end{equation}
would not impose such a constraint because $\hat{H}$ in general evolves $\ket{q_t}$ off $\mathcal{M}$.
An appealing resolution is given by the Time-Dependent Variational Principle (TDVP)~\cite{dirac1930note,haegeman2011time}, which proposes to first project the right-hand side onto the tangent space of the MPS manifold:
\begin{equation}
  \frac{d \ket{q_t}}{dt} = \hat{\mathcal{P}}_{T_{\mathcal{M}}} \hat{H} \ket{q_t}.
  \label{eq:tdvpeom}
\end{equation}
In this way, dynamics is restricted to the desired sub-space of MPS states~\cite{vanderstraeten2019tangent}.

It turns out that the time-discretized numerical integration of this $\ket{q_t}$ is amenable to efficient TN operations.
The technical details of the time-evolution algorithm have been reported elsewhere \cite{paeckel2019time,haegeman2011time,haegeman2016unifying,lubich2015time}.
We give only a brief, high-level overview of that procedure (see also App.~\ref{app:tdvp} for more details).
The central step in deriving the TDVP time-evolution algorithm is to demonstrate that the projector $\hat{\mathcal{P}}_{T_{\mathcal{M}}}$ can be decomposed in terms of a sum over $L$ terms (associated to each site of the MPS) and $L-1$ terms (associated to each bond connecting sites).
In the limit of a small time step $\delta t$, the propagator $e^{\hat{\mathcal{P}}_{T_{\mathcal{M}}} \hat{H} \delta t}$ can thus be evaluated as a composition of time-evolutions involving the $2L-1$ individual terms.
That composition of time-evolutions is executed by a one-site TDVP algorithm that ``sweeps'' from lattice site 1 through $L$, advancing each term for the small time step.

The efficiency of the TDVP time-evolution algorithm relies on the flexibility to represent the same MPS state in redundant ways due to a gauge freedom.
  Consider, for example, neighboring tensors $Q^{n_l}_{s_{l-1}s_l}$ and $Q^{n_{l+1}}_{s_ls_{l+1}}$ in an MPS.
  For an invertible $R$, replacing those tensors by $Q^{n_l}R$ and $R^{-1} Q^{n_{l+1}}$, respectively, does not change the MPS state; upon contracting over the neighboring sites, the identity $\mathbb{I} = R R^{-1}$ would fall out.
  By leveraging the gauge freedoms, it is always possible to transform Eq.~\eqref{eq:Pr_MPS} into a so-called mixed canonical form~\cite{vidal2003efficient}:
\begin{equation}
  \ket{q_t} = \hspace{-0.06in}\sum_{n_1, \hdots n_L} A^{n_1} \hdots A^{n_{l-1}} Q^{n_l} B^{n_{l+1}} \hdots B^{n_L} \ket{n_1, \hdots, n_L},
  \label{eq:mixedcanonical}
\end{equation}
where each $A$ is a left-orthogonal tensor satisfying $A^T A = \mathbb{I}$ and each $B$ is a right-orthogonal tensor satisfying $B B^T = \mathbb{I}$.
For notational compactness, we have suppressed implied summations over the bond indices between tensors in Eq.~\eqref{eq:mixedcanonical}.
We have retained the physical indices $n_1, \hdots, n_L$ and allow their presence to remind us that although tensors at each site are named merely $A$ and $B$, they are actually distinct tensors $A^{(1)}, A^{(2)}, \hdots, B^{(1)}, B^{(2)}, \hdots$ that vary across sites.
The mixed canonical form is said to be centered about the tensor $Q^{n_l}$.
From that tensor's perspective, the chain of $A$'s and chain of $B$'s constitute the environment, and it is beneficial to contract over the links between $A$'s and links between $B$'s to obtain so-called environment tensors.
These environment tensors inherit an orthogonality from the orthogonality of the $A$'s and $B$'s, significantly reducing the computational complexity to advance $Q^{n_l}$ in time by $\delta t$ provided the MPS was already centered about the $Q^{n_l}$ we aim to evolve.

\begin{figure*}[hbt]
\centering
\includegraphics[width=1\textwidth]{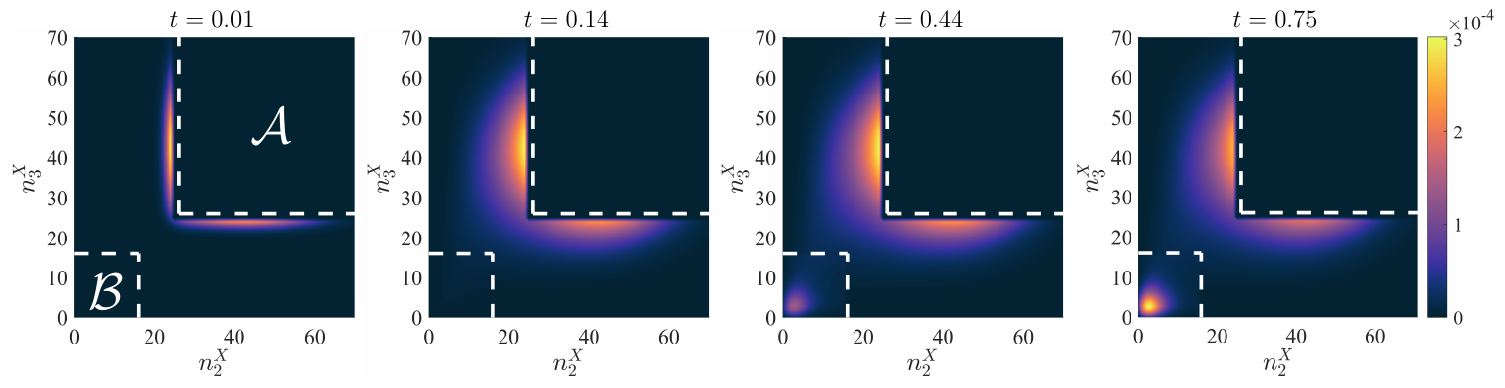}
\caption{\label{fig:Snaps} Snapshots of the joint density in two neighboring voxels from an $L=3$ site lattice of Schl\"{o}gl voxels.
  Using $\tilde{c}_1 = 2.676, \tilde{c}_2 = 0.040, \tilde{c}_3 = 108.102, \tilde{c}_4 = 37.881$ and $d = 8.2207$, the steady-state distribution was computed with $\kappa = 30$ by evolving an initial uniform distribution according to $\hat{H}$ via TDVP.
  That steady-state distribution was projected into $\mA$ at time 0, renormalized, then propagated according to the modified no-recrossing dynamics $\tilde{H}$ using TDVP.
  For clarity of visualization, the density within $\mA$ is not plotted as it would overwhelm the small probability that leaks into $\mB$.
  By $t = 0.01$ (left image), probability has begun to leak from $\mA$ but none has reached $\mB$.
  Probability has just begun to reach $\mB$ at at $t = 0.14$, roughly the $\tau_{\rm mol}$ timescale for this problem.
  By $t = 0.44$, probability is collecting in $\mB$, and by $t = 0.75$, the flux passing from $\mA$ to $\mB$ has attained an approximately constant steady-state value corresponding to the linear growth phase where $P(\mB_t|\mA) \propto k_{\mB\mA}t$.}
\end{figure*}

As made more explicit in App.~\ref{app:tdvp}, the one-site TDVP algorithm initially centers the MPS around the first tensor, propagates that tensor forward in time by $\delta t$, performs a QR decomposition to change the gauge, propagates the R term of that QR decomposition backward in time by $\delta t$, then passes that term to the neighboring site to re-center the MPS around the neighboring site.
The neighboring site is then propagated forward in time as the algorithm iterates.
After the complete sweep through the MPS, the probability distribution $\ket{q_t}$ has been propagated by a time step $\delta t$.
Whether that propagation well approximates the evolution of the full distribution $\ket{p_t}$ depends on the time-step error but also on the severity of the MPS approximation, which can be systematically controlled through the maximum bond dimension $\kappa$.

Notably, the time-evolved MPS state $\ket{q_t}$ allows us to estimate the probability of \emph{any} microstate of interest, even though it would be impossible to enumerate probabilities of all microstates.
Alternatively, sets of microstates can be summed over to yield tractable marginal distributions.
As an example, Fig.~\ref{fig:Snaps} shows the results of calculating $q_t(n_1, n_2, n_3)$ for a 1D chain of $L = 3$ Schl\"{o}gl sites.
Though there are three sites, site one is traced over to leave the two-dimensional joint probability of finding $n_2$ X molecules in site 2 and $n_3$ X molecules in site 3.
At time zero, the distribution is initialized within region $\mathcal{A}$ according to $\proj_\mA\ket{\boldsymbol{\pi}}/\bra{\mathbf{1}}\proj_{\mA}\ket{\boldsymbol{\pi}}$.
As time progresses, a small flux of probability leaks from $\mathcal{A}$ to $\mathcal{B}$.
Because the overwhelming preference is to stay in $\mA$, Fig.~\ref{fig:Snaps} highlights the rare transitions by only plotting the joint distribution outside of $\mA$.
Notice that, consistent with a steady-state approximation, while the population in $\mB$ gradually amasses, the distribution over microstates between $\mA$ and $\mB$ is nearly stationary.
The slight asymmetry between sites is due to the fact that site 2 is in the bulk while site 3 lies on the boundary.
The higher-dimensional joint $q_t(n_1, n_2, n_3)$ has the anticipated symmetry under interchange of $n_1$ and $n_3$.
More generally, symmetry holds between interchanging sites $l$ and $L-l+1$, as reflected in Fig.~\ref{fig:ExpMol}.
In the next section, we illustrate the fact that the MPS state's ability to capture the rare time-dependent flux is key to calculating the rate of traversing between basins.

\section{Many-body rate calculation}\label{sec:mbrates}
Having established the TDVP dynamics of an MPS state, we can now repeat the strategy of Eq.~\eqref{eq:rate} to compute a transition rate between states $\mathcal{A}$ and $\mathcal{B}$:
\begin{equation}
  \label{eq:tdvpdynamicsrate}
k_{\mathcal{B} \mathcal{A}} =  \frac{d}{dt} \frac{\bra{\mathbf{1}}\proj_{\mB}\,e^{\hat{\mathcal{P}}_{T_{\mathcal{M}}} \hat{H}t}\, \proj_{\mA} \ket{\boldsymbol{\pi}}}{\bra{\mathbf{1}}\proj_\mA\ket{\boldsymbol{\pi}}}\biggr\rvert_{t >\tau_{\rm mol}}.
\end{equation}
In doing so, there are a few important differences from Eq.~\eqref{eq:tdvpdynamicsrate}.
Most obviously, the TDVP dynamics includes the projection onto the MPS tangent space and $\ket{\boldsymbol{\pi}}$ must now be an MPS approximation to the steady-state distribution.
Additionally, the meaning of $\proj_{\mA}$ and $\proj_{\mB}$ must be adapted to now define many-body analogs of the projectors onto the two metastable states.
The single-voxel Schl\"{o}gl model had an $\mA$ state defined as all microstates with occupancy $n$ above a threshold  $\lambda_A^*$. Similarly, $\mB$ included all microstates with occupancy less than some other threshold $\lambda_B^*$.
The $L$-voxel Schl\"{o}gl model generalizes to involve hypercubic metastable states with basins defined by $\mA \equiv \{ n_l \geq \lambda_A^*\:\forall l\}$ and $\mB \equiv \{n_l \leq \lambda_B^*\; \forall l\}$.

With these generalizations, the transition rate of Eq.~\eqref{eq:tdvpdynamicsrate} would measure the flux crossing into $\mathcal{B}$ through its boundary, but if we are to compare with a transition path theory rate, we must exclude re-crossing events.
In other words, we want to include trajectories which exit $\mathcal{A}$ then some time later enter $\mathcal{B}$, but we want to exclude the flux that leaves $\mathcal{B}$ and re-enters $\mathcal{B}$ some time later without having returned to $\mathcal{A}$.
When $k_{\mathcal{BA}}$ is sufficiently large, those re-crossings can be ignored, but when $k_{\mathcal{BA}}$ becomes small enough, those recrossings can become significant and must be removed.
We note that the TDVP dynamics can just as well be executed with a modified absorbing-boundary-condition dynamics with effective Hamiltonian
\begin{equation}
  \label{eq:modified}
\tilde H = \tilde H^{\rm rxn} +\tilde H^{\rm diff},
\end{equation}
where the reaction and diffusion Hamiltonians exclude events that exit $\mB$.
An explicit construction of that modified absorbing-boundary dynamics is provided in App.~\ref{app:modifieddynamics}, allowing us to directly remove the re-crossing events.

\begin{figure*}[!ht]
\centering
\includegraphics[width=.99\textwidth]{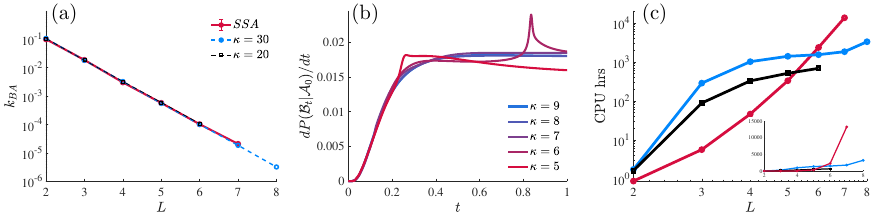}
\caption{\label{fig:RateL} (a) The rate $k_{\mB\mA}$ for switching between metastable states of the $L$-voxel Schl\"{o}gl model is calculated using the SSA (red line) and using TDVP (dashed lines) with varying bond-dimensions $\kappa$ using the same rate parameters as Fig.~\ref{fig:Snaps}.
  For sufficiently large bond-dimensions, TDVP agrees with a SSA rate estimate built using $400$ transitions from $\mA$ to $\mB$ at each $L$.
  (b) As $L$ grows a larger bond-dimension is required to capture the dynamics. Signatures of numerical instability are apparent when $\kappa$ is too small, shown here for $L = 3$, which stably converges when $\kappa$ exceeds 6.
  For this system, $\kappa=20$ is sufficient for numerical stability up to $L=6$, while $\kappa=30$ gives stable solutions up to $L=8$.
  (c) The total number of CPU hours to estimate $k_{\mB\mA}$ grows exponentially with SSA but polynomially with TDVP.}
\end{figure*}

\section{Numerical Evaluation of the $L$-Voxel Schl\"{o}gl Model Switching Rate}\label{sec:Results}
The formalism of the preceding sections has laid out a controllable approximation to extract rates between metastable states of RDMEs.
Numerical experiments are necessary to evaluate whether that formalism can be practically useful.
We set out to demonstrate that utility by computing the transition rates between high-occupancy and low-occupancy states of a $L$-voxel Schl\"{o}gl model as a function of $L$.
The model is parameterized by the four stochastic rate coefficients discussed in Sec.~\ref{sect:EstRate}.
We sought rate coefficients that fall in a bistable regime, with sharp peaks of steady-state probability falling in high- and low-occupancy states.
Since the dynamics would need to be truncated at some maximum occupancy $M$, we furthermore wanted to limit how many molecules would naturally accrue in the high-occupancy state.
By scanning parameters with the $L = 1$ Schl\"{o}gl model, we found parameters ($\tilde c_1 = 2.676, \tilde c_2 = 0.040, \tilde c_3 = 108.102,$ and $\tilde c_4 = 37.881$) compatable with an $M = 85$ truncation and with a sharp bistability between $\mA$ and $\mB$ regions defined by the thresholds $\lambda^*_{A} = 15$ and $\lambda^*_{B} = 25$.
By linking multiple voxels together in a 1D chain with a diffusive hopping rate $d$, it becomes rarer to switch between the two metastable states than it would be with a single isolated voxel.
With multiple voxels, a switching event must flip the state of each voxel, and diffusion between neighboring voxels works to stabilize neighbors in identical high- or low-occupancy states.
Due to that diffusive coupling between neighbors, the $L$-voxel Schl\"{o}gl model's $\mA$ to $\mB$ transition is characterized by a correlated flip of all voxels, and the rarity of that correlated event depends on $L$, the number of correlated voxels.
It is therefore straightforward for us to push the rate-calculation methodologies into challenging regimes; $k_{\mB \mA}$ can be decreased by growing $L$.

The most straightforward way to numerically detect the $L$-dependence of the transition rate is via sampling of trajectories.
Realizations of the continuous-time RDME, subject to the $M = 85$ maximum voxel occupancy at fixed $L$, were generated using the SSA.
First, $5\times 10^5$ independent SSA realizations were initialized from a uniform distribution and propagated in parallel up to $t=5$ to approach the steady state distribution.
Sampled configurations that fell within $\mathcal{A}$ were then sampled as initial configurations for 400 independent trajectories which were evolved via the SSA until the trajectory first reached $\mathcal{B}$.
From the $400$ trajectories, the transition rate was computed from the mean of those waiting times to reach $\mathcal{B}$: $k_{\mathcal{BA}} = \langle \tau_{\mathcal{BA}}\rangle^{-1}$.
Fig.~\ref{fig:RateL}a (red line) shows that the rate exponentially decreases as $L$ is increased, a result consistent with the exponential dependence of rate on volume in the well-mixed Schl\"{o}gl model~\cite{vlysidis2018differences}.
While the rate can be computed by trajectory sampling, the difficulty to do so (via brute force sampling) grows in proportion to the typical $\tau_{\mathcal{BA}}$, meaning the computational cost grows exponentially with $L$ (see Fig.~\ref{fig:RateL}c).

The poor scaling of brute-force SSA sampled rate calculations has motivated so-called advanced sampling methods like FFS \cite{allen2005sampling}, which focus computational effort on the progress along a reaction coordinate.
The idea is to introduce a reaction coordinate $\lambda$ that resolves progress in transitioning from a value $\lambda_A^*$ at the boundary of $\mA$ to $\lambda_B^*$ at the boundary of $\mB$.
Rather than seek long trajectories that snake from $\lambda_A^*$ all the way to $\lambda_B^*$, it can be beneficial to break the transition up into smaller trajectories that make some progress in moving to states with values of $\lambda$ closer to $\lambda_B^*$ than they started.
This FFS scheme is particularly clever in that it can be statistically unbiased, meaning any choice of reaction coordinate (any mapping taking a microstate to a value of $\lambda$) will yield the true rate in the infinite sampling limit.
Despite that formal lack of bias, the computational cost of a FFS calculation depends strongly on the correlation between $\lambda$ and the reaction's progress~\cite{allen2006simulating}.
If one already has a good idea of how the reactions rare transition events proceed, a smart $\lambda$ can be constructed and the exponential scaling of the SSA can be avoided, but that identification of the reaction coordinate is seldom straightforward.
For example, in the L-voxel Schl\"{o}gl model, will $\mA$ to $\mB$ events typically be triggered by a high- to low-occupancy flip at a boundary voxel of the 1D chain or in the bulk?
The boundary voxels have the benefit that they only have diffusion from one neighbor stabilizing the high-occupancy state, but the bulk voxels have the advantage of being more plentiful.
In App.~\ref{app:FFS} we describe FFS calculations employing two reasonable choices of reaction coordinate.
Those existing FFS methods can yield rates more cheaply than a brute force SSA calculation, but the speed up depends strongly on the reaction coordinate, the choice of which is nontrivial even for this simple model.

The DP/TDVP approach we have described offers a radically different strategy for circumventing the SSA scaling problem.
Because the reaction-coordinate selection problem is hard, we developed the DP/TDVP strategy to bypass it entirely, computing the rate and reaction mechanism without prior information about how the transition events move through the high-dimensional state space.
Fig.~\ref{fig:RateL} shows that the reaction-coordinate-agnostic DP/TDVP approach indeed reproduces the SSA rates at lower computational expense, reducing the scaling from exponential to polynomial in $L$.
The SSA rate calculations can be made more accurate by collecting more trajectories, and the DP/TDVP calculations likewise offer a systematic way to improve accuracy in exchange for a greater computational expense.
The two most prominent ways to adjust that accuracy are through the time step $\delta t$ for the TDVP evolution and through the maximum bond dimension $\kappa$ for the MPS state.
In this work, a time step of $\delta t = 10^{-4}$ was used throughout, and the trade-off between expense and accuracy was entirely tuned through the choice of $\kappa$.
The single-site TDVP algorithm we used to propagate an MPS state does not alter the bond dimension of the MPS, so $\kappa$ was entirely set by the bond dimension of the initial MPS.
We therefore needed to approximate the steady state $\ket{\boldsymbol{\pi}}$ with different values of $\kappa$.
Approximations were generated by starting with an initially uniform state using a bond dimension $\kappa$ set to several different values between 5 and 30.
Those uniform distributions were evolved with single-site TDVP for 10,000 time-steps allowing for the natural dynamics of $\hat{H}$ to approach a stationary state with $\bra{p_t}\ket{\dot{p}_t} = \bra{p_t}\hat{H}\ket{p_t} \approx 0$ by time $t = 1$. 
Irreducibility of $\hat H$ means the unique stationary state is the steady-state distribution $\ket{\boldsymbol{\pi}}$.
While $\bra{p_t}\hat{H}\ket{p_t} = 0$ would imply $\ket{p_t} = \ket{\boldsymbol{\pi}}$, $\bra{p_t}\hat{H}\ket{p_t} \approx 0$ does not guarantee that $\ket{p_t} \approx \ket{\boldsymbol{\pi}}$, only that the distribution is evolving very slowly.
When transitions between $\mA$ and $\mB$ are slow, one can imagine reaching a local steady state distribution within $\mA$ and  within $\mB$ but having probability very slowly repartitioning between the two.
For our purposes, what matters is not that $t = 1$ has exactly reached $\ket{\boldsymbol{\pi}}$ but rather that the $t = 1$ distribution confined to $\mA$, upon renormalization, has approached $\proj_\mA\ket{\boldsymbol{\pi}}/\bra{\mathbf{1}}\proj_{\mA}\ket{\boldsymbol{\pi}}$.
Provided dynamics within $\mA$ is comparatively fast, this local steady state within $\mA$ is reached more rapidly than $\ket{\boldsymbol{\pi}}$ and is reasonably confirmed by checking that $\bra{p_t}\hat{H}\ket{p_t} \approx 0$.

The renormalized $t = 1$ distribution was then propagated under single-site TDVP with the $\tilde{H}$ absorbing-boundary dynamics (see App.~\ref{app:modifieddynamics}) for another 10,000 time steps, and projected onto $\mathcal{B}$ to add up the total probability to have reached $\mathcal{B}$.
Note that these 10,000 steps are not nearly enough to reach the steady state $\ket{\boldsymbol{\pi}}$ because the slow rate of transitioning from $\mathcal{A}$ to $\mathcal{B}$ is a bottleneck.
 As additional time steps of dynamics were taken, the growth of the transition flux from $\mathcal{A}$ to $\mathcal{B}$ was measured and plotted in Fig.~\ref{fig:RateL}b.
That figure shows that at very short times ($t < \tau_{\rm mol}$), the flux from $\mathcal{A}$ to $\mathcal{B}$ has not reached a steady state value because there hasn't been enough time for any transitions.
After that initial time $\tau_{\rm mol}$, the plot plateaus at a value corresponding to the rate $k_{\mathcal{BA}}$, a plateau that is reached orders of magnitude faster than the time for a typical SSA trajectory to make a transition.
We find that if the bond dimension is insufficient (e.g., $\kappa = 5$ for $L = 3$ in Fig.~\ref{fig:RateL}b), then the dynamics becomes unstable, reflected in erratic estimates for the conditional probability $P(\mB_t|\mA_0)$.
As the bond dimension is grown (e.g., $\kappa = 9$ for $L = 3$ in Fig.~\ref{fig:RateL}b), the estimates converge, allowing a stable rate to be extracted.
That convergence as a function of $\kappa$ is analyzed further in App.~\ref{app:Err_kAB}.
We find that setting the bond dimension to roughly 10 or 20 is often sufficient, but larger values of $L$ can only converge with a larger bond dimension.
For example, Fig.~\ref{fig:RateL}a shows that SSA rates agree with $\kappa = 20$ calculations up to $L = 6$, but $\kappa = 30$ was required to push up to $L = 8$.
Provided that larger bond dimension is used, the DP/TDVP rates reproduce the SSA rates across five orders of magnitude, up to and beyond the point that the brute-force SSA rates are practical.

Though the required bond dimension grows with $L$, Fig.~\ref{fig:RateL}c shows that the expense to estimate the rate grows sub-exponentially.
The blue and black dashed lines in that plot are the total number of CPU hours time to estimate the rate using TDVP with $\kappa=30$ and $\kappa=20$ respectively.
Both SSA and TDVP were run using codes written in the Julia programming language \cite{bezanson2017julia} on equivalent CPUs with details of the runtime analysis described in App.~\ref{app:benchmarking}.
While the required CPU hours, of course, depends on details of the hardware, parallelization, the particular implementation, we illustrate the difference in scaling when comparing identical implementations run in identical ways across different system sizes.
Fig.~\ref{fig:RateL}c clearly demonstrates that difference in scaling between brute-force SSA and DP/TDVP.
An increase in bond dimension increases the computation time, but it does not fundamentally alter the favorable scaling in $L$.
Comparison of computational expense with FFS is complicated by the fact that one can optimize FFS in many different ways (how many interfaces, which reaction coordinate, etc.), but App.~\ref{app:FFS} shows that the DP/TDVP calculations remain favorable even compared to FFS with a reasonable guess of an order parameter.

\section{Deploying the tool}\label{sec:Deploying}
  We have described a numerical tool to extract the rate of a rare switching event from a reaction-diffusion model, but the numerical value of that rate is often not the final goal.
  More than the value of the rate, it is desirable to know how the rate changes in response to controllable parameters.
  Fig.~\ref{fig:RateL}a showed one such variation\textemdash the switching rate falls off exponentially with the number of voxels $L$.
  The essential physics of this exponential decay is consistent with a mechanism in which a boundary voxel of the 1D chain flips to create an interface between high- and low-occupation voxels.
  Subsequently, each voxel at the interface flips until the whole chain has flipped.
  Letting $p_{\rm end}$ denote the small rate of flipping a boundary voxel and $p_{\rm int}$ denote the small rate of flipping a voxel at the interface, it follows that at short times $P(\mB_t|\mA_0) \propto p_{\rm end} p_{\rm int}^{L-1}$.
  If $\left<\Delta t\right>$ is the typical time to wait between interface flips, one therefore expects the exponential decay with $k_{\mB \mA}(L) \propto e^{-L \ln [p_{int} \left<\Delta t\right>]}$.
  
\begin{figure}[t!]
\centering
\includegraphics[width=.4\textwidth]{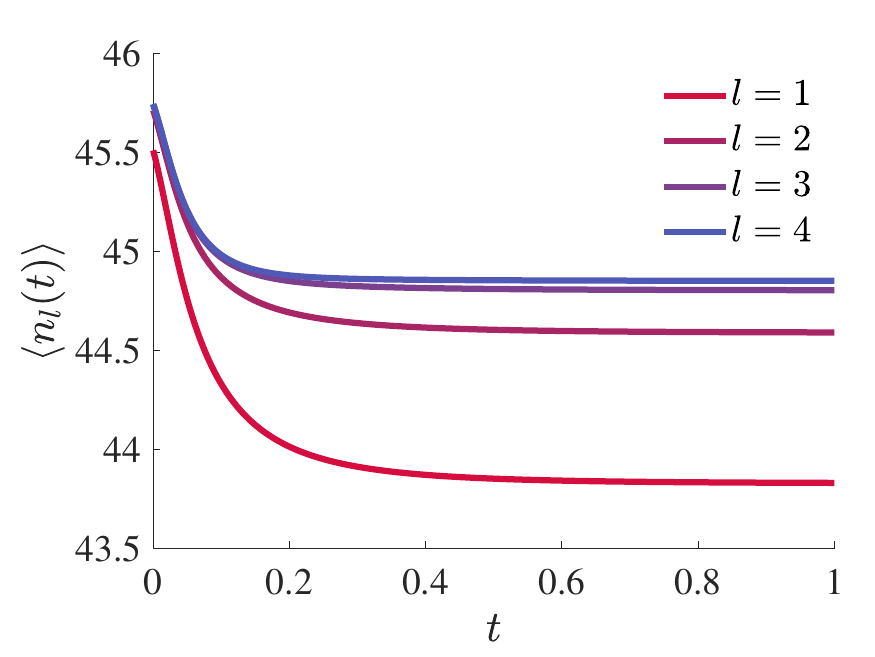}
\caption{\label{fig:ExpMol}
  The time evolution of the average number of molecules for different lattice sites when $L=8$ and rate parameters are as in Fig.~\ref{fig:Snaps}, computed with $\kappa = 30$.
  Different colors correspond to different voxels.
  The expected number of molecules $\langle n_l(t)\rangle = \bra{\mathbf{1}}\hat n_l\ket{p_t}$ for voxels $1$ to $\lfloor L/2\rfloor$ (shown) are mirrored by voxels $\lfloor L/2+1\rfloor$ to $L$.
  The initial non-linear flow of probability out of $\mA$ is shown by the steep drop in average number of molecules.
  The system passing $\tau_{mol}$ is shown by the transition to a linear change in particle number $\langle n_l(t)\rangle$.}
\end{figure}

The rate calculations as a function of $L$ thus hint at a stepwise mechanism with a flipped voxel passing from one end of the chain to the other, but that mechanistic insight can also be more directly extracted from the DP/TDVP evolution.
The time-evolution of $\ket{p_t}$ contains information about the entire distribution that flows from $\mA$ to $\mB$, and we can track the average number of molecules in each lattice site as a function of time.
For example, Fig.~\ref{fig:ExpMol} shows $\langle n_l(t)\rangle = \bra{\mathbf{1}}\hat n_l\ket{p_t}$, the average number of molecules in voxel $l$ with $l$ ranging from 1 through 4 for an $L = 8$ voxel lattice.
At time $t = 0$, the plot shows the normalized steady-state confined to $\mathcal{A}$, the high-occupation-number state.
In the steady state, the six bulk voxels, 2 through 7, have essentially indistinguishable mean occupancies, while the boundary voxels, 1 and 8, have slightly depressed means reflecting that it is marginally more likely for those two voxels to fluctuate to low-occupancy states.
As the time is allowed to evolve for a timescale $t < \tau_{\rm mol}$, the mean occupation of the boundary voxels drops markedly due to the influence of rare trajectories in which one or both boundary voxels can flip.
Subsequent to that drop in $\left<n_1(t)\right>$, the mean occupancies of voxels 2 then 3 also drop, though less sharply, establishing the mechanistic sequence.
After the molecular timescale $\tau_{\rm mol}$, the rate of decrease in $\langle n_l(t)\rangle$ becomes linear, reflecting the steady-state flux from $\mathcal{A}$ to $\mathcal{B}$.

\begin{figure}[t!]
\centering
\includegraphics[width=.4\textwidth]{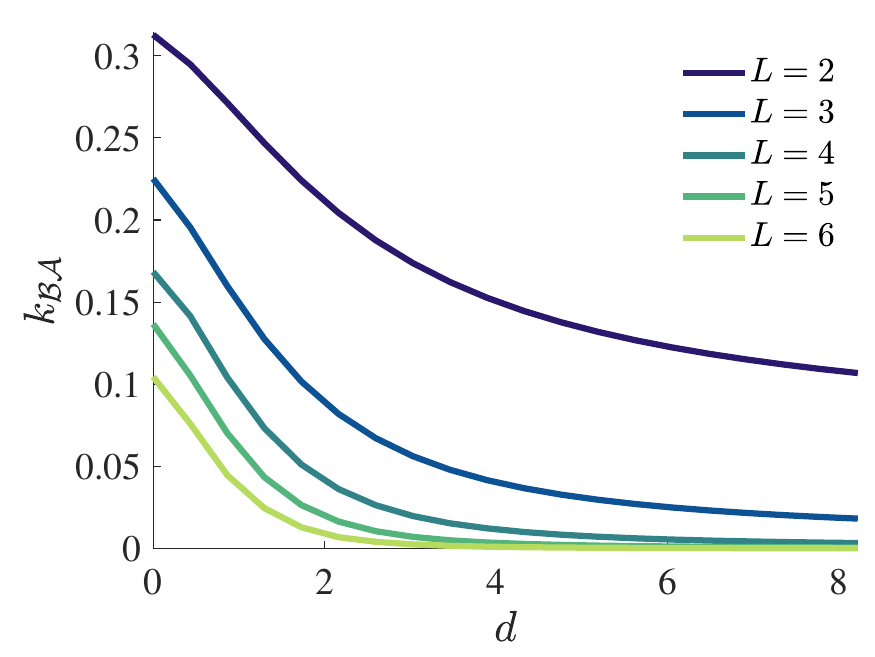}
\caption{\label{fig:VaryD} The switching rate $k_{\mB\mA}$ as a function of diffusion coefficient $d$ for $2\leq L \leq 6$, computed with $\kappa = 20$ for the rates from Fig.~\ref{fig:Snaps}.
  As $d$ is increased, the diffusion acts to inhibit the stochastic switching of each voxel.
  At large enough $d$ the system acts similar to a large well-mixed system, resulting in a leveling off of the rate coefficients.
}
\end{figure}

In addition to varying the extent of the space, it is interesting to tune the balance between reaction and diffusion events through the diffusive hopping rate $d$.
The sequential mechanism suggests that a limitation to the transition events emerges from $p_{\rm int}$, the probability of flipping the state of an interior voxel at an interface between low- and high-occupation states.
The transition events thus have low- and high-occupation blocks separated by an interface that diffuses until the low-occupation block covers the entire system.
When molecules diffuse between interface voxels, the interface itself can diffuse, and the faster the interface sweeps through the system, the faster the transition events.
One might therefore anticipate that the flipping rate grows with $d$.
In fact, we find the exact opposite.
Fig.~\ref{fig:VaryD} shows the results from varying the diffusion over 20 evenly spaced values while holding the bond-dimension fixed at $\kappa = 20$. The rate decreases for each value of $L$ as $d$ is increased due to diffusion acting as an effective stabilizer of high-occupation states.

At $d = 0$, each voxel is an independent well-mixed system.
The switching rate depends on the probability that each voxel independently experiences a collapse of population, dislodging it from its metastable high-occupation state.
As $d$ grows, there is a diffusive mechanism to fight against that population collapse.
When the population of a voxel wanders dangerously close to a tipping point, molecules could diffuse from a high-occupancy neighbor to restore the population.
In the limit of very fast diffusion, the entire chain of voxels effectively behaves as a single large well-mixed system with a markedly lower switching rate because statistical fluctuations away from the metastable state get correspondingly smaller.

\section{Discussion}
\label{sec:Discussion}
Using the 1D Schl\"{o}gl model, we have demonstrated how TN calculations combined with the DP framework offer a new paradigm for numerical investigation of reaction-diffusion models, even when the physical bond dimension ($M+1$) is not small.
That demonstration opens up a large class of RDME models which represent small well-mixed reactors, connected together by diffusive coupling.
As illustrated in this work, these models can exhibit interesting physics when each reactor is small enough to exhibit significant statistical fluctuations and when the diffusion between reactors induces correlations between fluctuations in different reactors.
In this work, our approach was to start with a well-studied kinetic toy model and to add spatial dynamics.
Spatial extensions of other toy models like the Brusselator~\cite{tyson1973some} or the Oregonator~\cite{field1974oscillations} would be amenable to similar numerical analysis because any reaction-diffusion model which can be written as a set of elementary reactions will have a DP representation.
That representation can involve creation and annihilation operators for different particle types, not only for species X, but the mapping from reactions to effective Hamiltonians is systematic.
Table~\ref{tbl:Rxn} shows the effective Hamiltonians for several common (unchemostatted) reactions: unimolecular, bimolecular, and auto-catalytic.
Given the set of reactions, it is comparatively straightforward to derive the DP representation, and computational tools already exist to generate an MPO~\cite{schollwock2011density} and propagate via TDVP from that DP form of the effective Hamiltonian.
\begin{table}[ht]
\centering
\begin{tabular}{c|c}
Reaction                 & Hamiltonian                                                       \\ \hline
\ce{X ->[$\tilde{c}$] Y}        & $\tilde{c}[(y^\dagger - x^\dagger) x]$                           \\[1pt]
\ce{X +Y ->[$\tilde{c}$] Z}     & $\tilde{c}[(z^\dagger - x^\dagger y^\dagger)x y]$                \\[1pt]\ce{X + Y ->[$\tilde{c}$] 2X} & $\tilde{c}[x^{\dagger }(x^\dagger - y^\dagger)x y]$ \\[1pt] \hline
\end{tabular}
\caption{\label{tbl:Rxn}
  Examples of the mapping between a common chemical reaction and the Doi-Peliti form of the contribution to an effective Hamiltonian.
  The procedure to derive such correspondences is systematic~\cite{schulz2005exact} and typically discussed for a Fock space with occupation number ranging from 0 to infinity.
  To truncate to a finite Fock space, one must additionally add a projector to each effective Hamiltonian, as in this work, to ensure conservation of probability.
  }
\end{table}

As the TN techniques become increasingly well developed, one can imagine that the computational tools will offer generic ``turn the crank'' analysis of microscopic models of reaction-diffusion at the ensemble level.
We anticipate these numerical approaches to be beneficial to both synthetic and biophysical reaction-diffusion systems.
In the case of the former, the discretization of space into reactors can be explicitly realized, as in the microfluidic fabrication of a two-dimensional array of nanoliter-volume Belousov–Zhabotinsky (BZ) reactors~\cite{litschel2018engineering}.
In line with our 1D Schl\"{o}gl model, the strength of coupling between those BZ nanoreactors is regulated by diffusion, and there is experimental interest in pattern formation as a function of that diffusive coupling.
In the case of biophysics, sub-cellular dynamics is dominated by the generation and diffusion of proteins with small copy numbers.
Much work has been devoted to understanding the logic of cellular decision making through stochastic CRN models, most famously gene regulatory models, which are most frequently studied in a well-mixed limit~\cite{warren2004enhancement}.
We see these TN tools as an intriguing pathway to numerically investigate the timescales and pathways for switching between metastable states in well-mixed CRNs like gene toggle switches~\cite{allen2005sampling} and in reaction-diffusion variants of those gene-regulatory models.
Eventually, one can dream of bringing these methods to the broad array of systems biology applications where RDME models are widely simulated~\cite{wilkinson2009stochastic}, including in large-scale GPU-accelerated simulations of systems with biologically-relevant size~\cite{hallock2014simulation}.
At present those sorts of reaction-diffusion model can only be numerically accessed through sampling approaches, and we expect the ensemble perspective will some day complement those efforts.

To move from the toy models to the synthetic and biophysical systems, it will be important to expand the ability of TNs to approximate correlation between species and voxels that are not simply connected through a 1D chain.
Here, because our 1D chain of Schl\"{o}gl reactors are connected only to their neighbors, it was natural to approximate the microscopic distribution with an MPS.
That choice allowed us to capitalize on the efficient TDVP algorithm for MPS states, but a RDME with many dynamic species will not always be easily approximated by a 1D chain of tensors.
If two chemical species are strongly correlated but far apart in the MPS structure, the MPS calculation will demand a very high bond dimension.
It will be important for future work to dynamically adapt the bond dimension through subspace expansion~\cite{hubig2015strictly} or by devising systematic ways to generate tractable TN architectures with connectivity mirroring the correlations between CRN species.
Those technical advances may eventually lay the groundwork to move the methodology from this Schl\"{o}gl study to the complex, biologically-motivated models where SSA realizations presently reign supreme.
 
\section{Acknowledgments}
We gratefully acknowledge Hadrien Vroylandt, Nils Strand, Geyao Gu, and Luis Pedro Garcia-Pintos for many insightful discussions.
We are also grateful to Miles Stoudenmire, Matthew Fishman, Steven White, and other developers of ITensor, a library for implementing tensor network calculations, upon which this work was built.
This research was supported in part through the computational resources and staff contributions provided for the Quest high performance computing facility at Northwestern University which is jointly supported by the Office of the Provost, the Office for Research, and Northwestern University Information Technology.
  The material presented in this manuscript is based upon work supported by the National Science Foundation under Grant No.\ 2141385.
  TRG is thankful that Phill Geissler could see this work in a preliminary form, but is deeply saddened that Phill's much-too-soon passing prevented him from sharing more of his wisdom.
  
 \appendix
  \section{Doi-Peliti encoding}\label{app:DPEnc}
  Assume a set of elementary reactions are given.
  These can be encoded into a Doi-Peliti representation which converts each elementary reaction into a term of an effective Hamiltonian.
  The operator representations can then be exactly written as an MPO, as shown below.
  Consider an $L$-site $1$D lattice of well-mixed voxels.
  Inside each voxel the Schl\"ogl model reactions can fire.
  A probability distribution will be written as a vector in the Fock space $\mathcal{F}^L$ spanned by orthonormal basis vectors $\ket{\bfn} = \ket{n_1,n_2,\ldots,n_L}$ which correspond to single microstates, with occupancy of each voxel given by $n_1, n_2, \hdots, n_L$.
  For any two microstates $\ket{\bfn}$ and $\ket{\mathbf{m}}$ we thus have $\bra{\mathbf{m}}\ket{\bfn} = \delta_{\mathbf{m},\bfn}$.
  Distributions over the microstates can be written as $\ket{p_t} = \sum_{\bfn} p_t(\bfn) \ket{\bfn}$, where $p_t(\bfn) \equiv p(\bfn, t) \equiv p(n_1, n_2, \hdots, n_L, t)$ is the probability of microstate $\bfn$ at time $t$.
  When focusing on the $l^{\rm th}$ voxel we will use the shorthand $p_t(\bfn) \equiv p_t(\bfn^c, n_l)$, where $\bfn^c = \left\{n_k, \forall k \neq l\right\}$ defines the state of the set of voxels that complement voxel $l$.
  Each reaction of the Schl\"ogl model contributes to the set of differential equations which describe the time evolution of $p_t(\bfn)$. We must convert the local change in probability of microstate $\bfn$ to the global action of the effective Hamiltonian on $\ket{p_t}$. The linear nature of the chemical master equation means we can consider one reaction at a time.

  \subsection{Reaction 1: \ce{$2$X + A ->[\tilde{c}_1] $3$X}}
  We start with the reaction $\ce{$2$X + A ->[\tilde{c}_1] $3$X}$.
  When this reaction fires inside site $l$, the change in probability is
  \begin{align}
    \frac{dp_t(\bfn)}{dt} = \frac{\tilde c_1}{2}  & \left[(n_l-1)(n_l-2)p_t(\bfn^c,n_l-1)\right. \nonumber \\
      & \ \ \ \ - \left.n_l(n_l-1)p_t(\bfn^c,n_l)\right].
    \label{eq:rxn1init}
  \end{align}
  The first (positive) term captures the gain of probability into microstate $\bfn$ from microstate $(\bfn^c, n_l - 1)$.
  Two of the $n_l - 1$ possible X molecules must come together to make the third, and there are $(n_l - 1)(n_l - 2) / 2$ total combinations. The second (negative) term of Eq.~\eqref{eq:rxn1init} accounts for the probability of already being in state $\bfn$ at time $t$. Notice that this reaction adds an X molecule, so if the system is already at the maximum cap of $n_l = M$ molecules, there can be no negative term at $M+1$, to balance out conservation of probability.
  The loss term therefore requires extra care when there is an enforced cap since the loss term would come from a forbidden transition.
Consequently, Eq.~\eqref{eq:rxn1init} can only apply to microstates $\bfn$ whose $l^{\rm th}$ voxel is not maximally occupied.
  For the general case, we could rewrite the evolution as
    \begin{align}
    \frac{dp_t(\bfn)}{dt} = \frac{\tilde c_1}{2}  & \left[(n_l-1)(n_l-2)p_t(\bfn^c,n_l-1)\right. \nonumber \\
      & \ \ \ \ - \left. (1 - \delta_{n_l, M}) n_l(n_l-1)p_t(\bfn^c,n_l)\right].
    \label{eq:rxn1init2}
    \end{align}
    Summing each side of Eq.~\eqref{eq:rxn1init2} over the Fock states $\ket{\bfn}$ yields
    \begin{align}
      \frac{d\ket{p_t}}{dt} = \frac{\tilde c_1}{2}\sum_{\bfn} & \left[(n_l-1)(n_l-2)p_t(\bfn^c,n_l-1)\ket{\bfn}\right. \nonumber \\
        & \ \ \ \ - \left. (1 - \delta_{n_l, M}) n_l(n_l-1)p_t(\bfn^c,n_l)\ket{\bfn}\right].
      \label{eq:rxn1fock}
    \end{align}
    It is useful to match the argument of $p_t$ with the ket, which can be achieved by introducing the appropriate creation operator:
    \begin{align}
      \nonumber & \frac{d\ket{p_t}}{dt}\\
      & \ = \frac{\tilde c_1}{2}\sum_{\bfn} \bigg[ \left(n_l-1\right)\left(n_l-2\right)p_t(\bfn^c,n_l-1)x_l^\dagger \ket{\bfn^c, n_l-1} \nonumber \\
        & \ \ \ \ \ \ \ \ \ \ \ \ \ \ \ -  (1 - \delta_{n_l, M}) n_l(n_l-1)p_t(\bfn^c,n_l)\ket{\bfn}\bigg].
      \label{eq:rxn1fock2}
    \end{align}
    The $x^\dagger_l$ term preceding the first ket is an operator, but all other terms preceding the kets are numbers that readily commute.
    We reorder those terms to get
    \begin{align}
      \nonumber & \frac{d\ket{p_t}}{dt}\\
      & \ = \frac{\tilde c_1}{2}\sum_{\bfn} \bigg[x_l^\dagger p_t(\bfn^c,n_l-1) \left(n_l-1\right)\left(n_l-2\right) \ket{\bfn^c, n_l-1} \nonumber \\
        & \ \ \ \ \ \ \ \ \ \ \ \ \ \ \ -  (1 - \delta_{n_l, M}) p_t(\bfn^c,n_l) n_l(n_l-1) \ket{\bfn}\bigg].
      \label{eq:rxn1fock3}
    \end{align}
    Eqs.~\eqref{eq:raiselower} give the identity $x_l^{\dagger 2} x_l^2 \ket{\bfn} = n_l(n_l - 1) \ket{\bfn}$, allowing both the gain and loss terms to be re-expressed in terms of powers of the creation and annihilation operators:
    \begin{align}
      \frac{d\ket{p_t}}{dt} = \frac{\tilde c_1}{2}\sum_{\bfn} &\bigg[x_l^\dagger p_t(\bfn^c,n_l-1) x_l^{\dagger 2} x_l^2 \ket{\bfn^c, n_l-1} \nonumber \\
        & \ \ -  (1 - \delta_{n_l, M}) p_t(\bfn^c,n_l) x_l^{\dagger 2} x_l^2 \ket{\bfn}\bigg].
      \label{eq:rxn1fock4}
    \end{align}
    Notice that the creation and annihilation operators can now be factorized outside of the sum to yield two terms which look very similar to the Fock-space expansion $\ket{p_t} = \sum_{\bfn} p_t(\bfn)\ket{\bfn}$.
    \begin{align}
      \frac{d\ket{p_t}}{dt} = \frac{\tilde c_1}{2} &\bigg[x_l^{\dagger 3} x_l^2 \sum_{\bfn} p_t(\bfn^c,n_l-1) \ket{\bfn^c, n_l-1} \nonumber \\
        & \ \ -  x_l^{\dagger 2} x_l^2 \sum_{\bfn} (1 - \delta_{n_l, M}) p_t(\bfn^c,n_l) \ket{\bfn}\bigg].
      \label{eq:rxn1fock5}
    \end{align}
    The first term is like the expansion of $\ket{p_t}$ except that the bounds of summation are shifted, including kets with occupancy $-1$ through $M_l - 1$ rather than $0$ through $M_l$.
    The non-physical state with occupancy $-1$ has no probability, so it can be excluded from the sum.
    Furthermore, the occupancy $M$ term can be added to the sum since $x_l^{\dagger 3} x_l^2 \ket{\bfn^c, M_l} = 0$.
    Consequently, the first sum of Eq.~\eqref{eq:rxn1fock5} is equivalent to $\ket{p_t}$.
    The second sum can also be expressed in terms of $\ket{p_t}$ if we recognize that the delta function constraint can be viewed as the eigenvalue of a projector operator:
    \begin{equation}
      \proj_{M_l}^c \ket{\bfn} = \bigg(\mathbb{I} - \ket{\bfn^c, M_l}\bra{\bfn^c, M_l}\bigg) \ket{\bfn} = \left(1 - \delta_{n_l, M}\right)\ket{\bfn}.
    \end{equation}
    Hence, both sums of Eq.~\eqref{eq:rxn1fock5} simplify in terms of the same Fock state $\ket{p_t}$:
    \begin{equation}
      \frac{d\ket{p_t}}{dt} = \frac{\tilde c_1}{2} \left[x_l^{\dagger 3} x_l^2  - x_l^{\dagger 2} x_l^2 \proj^c_{M_l}\right]\ket{p_t}.
      \label{eq:rxn1fock6}
    \end{equation}
    The factor of $1/2$ can be traced back to a combinatorial counting term for the unique number of ways to label the two X molecules that react.
    We absorb that counting term into the rate $c_1 = \tilde{c_1}/2$ and extract from Eq.~\eqref{eq:rxn1fock6} the contribution to the effective Hamiltonian arising from the first reaction in site $l$:
    \begin{equation}
      \hat{H}_l^{A_{\rm for}} = c_1\left[ x_l^{\dagger 3}x_l^2 - x_l^{\dagger 2}x_l^2 \proj^c_{M_l}\right].
      \label{eq:rxn1for}
    \end{equation}
    We take that direction of the reaction to be ``forward'' and use the superscript A because this is the reaction mediated by species A.
    By applying Eq.~\eqref{eq:realcomm}, the projector term can be rewritten in terms of a commutator of $x_l$ and $x_l^\dagger$ to give an alternative form of $\hat{H}_l^{A_{\rm for}}$ expressed only in terms of the creation and annihilation operators, $c_1$, and the maximal voxel occupancy $M$:
    \begin{equation}
      \hat{H}_l^{A_{\rm for}} = c_1\left[ x_l^{\dagger 3}x_l^2 - \frac{x_l^{\dagger 2}x_l^3 x_l^\dagger}{M+1} + \frac{x_l^{\dagger 2}x_l^2 x_l^\dagger x_l}{M+1} - \frac{M x_l^{\dagger 2}x_l^2}{M+1}\right].
      \label{eq:rxn1foralt}
    \end{equation}
    This form, shows that in the limit $M \to \infty$ Eq. \ref{eq:rxn1foralt} recovers the more standard bosonic result, $\hat{H}_l^{A_{\rm for}} = c_1 \left[x_l^{\dagger 3}x_l^2 - x_l^{\dagger 2}x_l^2\right]$.
    For compactness, we typically leave our expressions for contributions to the effective Hamiltonians as expressions that explicitly involve the finite-voxel projector, as in Eq.~\eqref{eq:rxn1for}.
    
    \subsection{Reaction 2: \ce{$3$X ->[\tilde{c}_2] $2$X + A}}
    The reverse reaction, which transitions from an occupancy of $n_l + 1$ to $n_l$, has $(n_l + 1) n_l (n_l-1) / 6$ different ways to combine three of the X molecules together.
    Consequently, the change in probability for the reverse reaction is
    \begin{align}
      \frac{dp_t(\bfn)}{dt} = \frac{\tilde c_2}{6}  & \left[(n_l-1)n_l(n_l+1) p_t(\bfn^c,n_l+1)\right. \nonumber \\
        & \ \ \ \ - \left.n_l(n_l-1)(n_l-2) p_t(\bfn^c,n_l)\right].
      \label{eq:rxn2init}
    \end{align}
    Unlike the forward reaction, the negative term does not include a factor of $\delta_{n_l, M}$ that treats the maximal occupancy state differently than the others.
    The distinction is that the loss terms from reaction 2 arise when probability is lost from occupancy $n_l$ down to occupancy $n_l - 1$.
    Those loss terms are possible from $n_l = 3$ through $M$.
    If there are fewer than 3 molecules, the reaction 2 mechanism is not possible, but this restriction emerges naturally from the $n_l(n_l-1)(n_l-2)$ coefficients in Eq.~\eqref{eq:rxn2init}. 
    Expanding in the Fock basis therefore gives
    \begin{align}
      \frac{d \ket{p_t}}{dt} = \frac{\tilde c_2}{6} \sum_{\bfn} & \left[(n_l-1)n_l(n_l+1) p_t(\bfn^c, n_l+1) \ket{\bfn}\right. \nonumber \\
        & \ \ \ \ -\left. n_l(n_l-1)(n_l-2) p_t(\bfn) \ket{\bfn}\right].
      \label{eq:rxn2fock}
    \end{align}
    Mirroring the treatment of the forward reaction, we rewrite the first $\ket{\bfn}$ using a lowering operator
    \begin{align}
      \frac{d \ket{p_t}}{dt} = \frac{\tilde c_2}{6} \sum_{\bfn} & \left[(n_l-1)n_l p_t(\bfn^c, n_l+1) x_l \ket{\bfn^c, n_l+1}\right. \nonumber \\
        & \ \ \ \ -\left. n_l(n_l-1)(n_l-2) p_t(\bfn) \ket{\bfn}\right].
      \label{eq:rxn2fock2}
    \end{align}
    Eqs.~\ref{eq:raiselower} imply $x^{\dagger 2}x_l^3 \ket{\bfn^c, n_l + 1} = (n_l - 1)n_l x_l \ket{\bfn^c, n_l+1}$ and $x^{\dagger 3} x_l^3 \ket{\bfn} = n_l(n_l-1)(n_l-2) \ket{\bfn}$, which allow us to simplify Eq.~\eqref{eq:rxn2fock2} to
    \begin{align}
      \frac{d \ket{p_t}}{dt} = \frac{\tilde c_2}{6} & \left[x^{\dagger 2}x_l^3 \sum_{\bfn} p_t(\bfn^c, n_l+1) \ket{\bfn^c, n_l+1}\right. \nonumber\\
        & \ \ \ \ -\left. x^{\dagger 3} x_l^3 \sum_{\bfn} p_t(\bfn) \ket{\bfn}\right].
      \label{eq:rxn2fock3}
    \end{align}
    We recognize the second sum over $\bfn$ as giving $\ket{p_t}$.
    The first sum ranges from $n_l = 1$ through $M+1$, but the $M+1$ term vanishes due to the occupancy cap.
    Furthermore, an $n_l = 0$ term can be added to the sum since that term also vanishes ($x^{\dagger 2}_l x_l^3 \ket{\bfn^c, 0_l} = 0$).
    Thus the first sum can be re-indexed to give $\ket{p_t}$.
    Performing both of those sums yields
    \begin{equation}
      \frac{d\ket{p_t}}{dt} = \frac{\tilde c_2}{6} \left[x_l^{\dagger 2} x_l^3  - x_l^{\dagger 3} x_l^3 \right]\ket{p_t}.
      \label{eq:rxn2fock4}
    \end{equation}
    The contribution to the effective Hamiltonian from the reverse A-mediated reaction in voxel $l$ is thus
    \begin{equation}
      \hat{H}_l^{A_{\rm rev}} = c_2\left[ x_l^{\dagger 2}x_l^3 - x_l^{\dagger 3}x_l^3 \right],
      \label{eq:rxn1rev}
    \end{equation}
    where we have again absorbed the combinatorial term into $c_2 = \tilde{c_2}/6$.
    By summing together the forward and reverse reactions, we arrive at the effective Hamiltonian associated with the reversible $A$-mediated reaction, quoted in Eq.~\eqref{eq:HA}:
    \begin{align}
      \nonumber \hat{H}_l^A &= \hat{H}_l^{A_{\rm for}} + \hat{H}_l^{A_{\rm rev}}\\
      &= c_1\left[ x_l^{\dagger 3}x_l^2 - x_l^{\dagger 2}x_l^2 \proj^c_{M_l}\right] + c_2\left[ x_l^{\dagger 2}x_l^3 - x_l^{\dagger 3}x_l^3 \right]
    \end{align}

    \subsection{Reaction 3: \ce{B ->[\tilde{c}_3] X}}
    Deriving the effective Hamiltonian for the third reaction is even more straightforward than the previous two.
    As in the first reaction, the loss term is only present if the microstate is not maximally occupied, so the probability for microstate $\bfn$ evolves according to
\begin{equation}
  \frac{dp_t(\bfn)}{dt} = \tilde{c}_3p_t(\bfn^c,n_l-1) - \left(1 - \delta_{n_l, M}\right)\tilde{c}_3 p_t(\bfn).
  \label{eq:rxn3}
\end{equation}
In the Fock basis, this evolution becomes
\begin{equation}
  \frac{d \ket{p_t}}{dt} = \tilde{c}_3 \sum_{\bfn} \left[p_t(\bfn^c, n_l-1) \ket{\bfn} - (1 - \delta_{n_l, M}) p_t(\bfn) \ket{\bfn}\right].
  \label{eq:rxn3fock}
\end{equation}
We again seek to make the first ket's microstate match the argument of $p_t$ by introducing the creation operator:
\begin{align}
  \nonumber \frac{d \ket{p_t}}{dt} &= \tilde{c}_3 \sum_{\bfn} \bigg[p_t(\bfn^c, n_l-1) x_l^\dagger \ket{\bfn^c, n_l-1} \nonumber \\
    \nonumber  & \ \ \ \ \ \ \ \ \ - (1 - \delta_{n_l, M}) p_t(\bfn) \ket{\bfn}\bigg]\\
    \nonumber &= \tilde{c}_3 \bigg[ x_l^\dagger \sum_{\bfn} p_t(\bfn^c, n_l-1) \ket{\bfn^c, n_l-1} \nonumber \\
      & \ \ \ \ \ \ \ \ \ - \sum_{\bfn} (1 - \delta_{n_l, M}) p_t(\bfn) \ket{\bfn}\bigg].
  \label{eq:rxn3fock2}
\end{align}
As in reaction 1, we re-index the first sum to get $\ket{p_t}$ and we introduce the projector operator into the second sum to get $\proj^c_{M_l} \ket{p_t}$, leaving
\begin{equation}
  \frac{d \ket{p_t}}{dt} = \tilde{c}_3 \left(x_l^\dagger - \proj^c_{M_l}\right) \ket{p_t}
  \label{eq:rxn3fock3}
\end{equation}
In this case, there is no combinatorial counting term, so $c_3 = \tilde{c}_3$ and
\begin{equation}
  \hat{H}_l^{B_{\rm for}} = c_3 \left(x_l^\dagger - \proj^c_{M_l}\right).
  \label{eq:rxn3final}
\end{equation}

\subsection{Reaction 4: \ce{X ->[\tilde{c}_4] B}}
The reverse reaction is like reaction 2 in that the reaction decreases X and therefore cannot cause an overflow beyond the maximal voxel occupancy.
Consequently, the evolution of probability in microstate $\bfn$ is
\begin{equation}
\frac{dp_t(\bfn)}{dt} = \tilde{c}_4(n_l+1)p_t(\bfn^c,n_l+1) - \tilde{c}_4n_lp_t(\bfn^c,n_l).
\end{equation}
Expanding in the Fock space gives
\begin{align}
\nonumber  \frac{d\ket{p_t}}{dt} &= \tilde{c}_4 \sum_{\bfn} \bigg[(n_l+1) p_t(\bfn^c, n_l+1) \ket{\bfn} - n p_t(\bfn) \ket{\bfn}\bigg]\\
  &= \tilde{c}_4 \sum_{\bfn} \bigg[p_t(\bfn^c, n_l+1) x_l \ket{\bfn^c, n_l+1} - n p_t(\bfn) \ket{\bfn}\bigg].
\end{align}
As before, we expand into two sums.
Upon checking the two boundary terms of the first sum, we re-index it and sum over microstates, recovering $\ket{p_t}$ from both sums to get
\begin{equation}
  \frac{d \ket{p_t}}{dt} = \tilde{c}_4 \left(x_l - x_l^\dagger x_l\right) \ket{p_t}.
\end{equation}
The combinatorial term is again trivial, so $\tilde{c}_4 = c_4$, and
\begin{equation}
  \hat{H}_l^{B_{\rm rev}} = c_4 \left(x_l - x_l^\dagger x_l\right).
  \label{eq:rxn4final}
\end{equation}
Combining Eqs.~\eqref{eq:rxn3final} and~\eqref{eq:rxn4final} yields the effective Hamiltonian arising from the reversible $B$-mediated reaction, Eq.~\eqref{eq:HB}:
\begin{equation}
  \hat{H}_l^B = c_3 \left(x_l^\dagger - \proj^c_{M_l}\right) + c_4 \left(x_l - x_l^\dagger x_l\right).
\end{equation}
The diffusion operator $\hat{H}^{\rm diff}$ of Eq.~\eqref{eq:Hdiff} is derived similarly by adding the maximal voxel occupancy constraint to the derivation of the bosonic diffusion operator~\cite{cardy2006}.

\section{Matrix Product Operators}\label{app:MPO}
Notice that the terms contributing to the effective Hamiltonian (Eqs.~\eqref{eq:HA},~\eqref{eq:HB}, and~\eqref{eq:Hdiff}) each take a ``product'' form in which the creation and annihilation operators at a site multiply other creation and annihilation operators at that same site or at a nearest-neighbor site.
For systems with local and nearest-neighbor interactions, there exists a relatively simple representation of the effective Hamiltonian.
Let $\hat{H}_l = \hat{H}_l^A + \hat{H}_l^B$ be the local effective Hamiltonian of the reversible reactions occurring in voxel $l$.
The total effective Hamiltonian $\hat{H}$ includes this local contribution summed over all voxels $l$ plus the $\hat{H}^{\rm diff}$.
Owing to the nearest-neighbor interactions, these terms can be elegantly summed via a product of operator-valued vectors and matrices.
We define the operator-valued vectors associated with voxels $1$ and $L$ as
\begin{equation}
W^{[1]} = \left[ \hat{H}_1, d x_1, d\mathcal{\hat{P}}^c_{M_1}, d x_1^\dagger, -x_1^\dagger x_1, \mathbb{I}\right], \;\;\; 
W^{[L]}= \begin{bmatrix}\mathbb{I} \\  x_{L}^\dagger \\ -x_L^\dagger x_{L}\\ x_L \\ d\mathcal{\hat{P}}^c_{M_L}\\ \hat{H}_L\end{bmatrix}.
\end{equation}
and the operator-valued matrix associated with voxels $2$ through $L-1$ as
\begin{equation}
W^{[l]} = \begin{bmatrix}
\mathbb{I} &  & & & & \\
x_{l}^\dagger & & & & & \\
-x_{l}^\dagger x_l & & & \bigzero & & \\
x_l     & & & & & \\
d\mathcal{\hat{P}}^c_{M_l}     & & & & & \\
\hat{H}_l, & d x_{l},  & d\mathcal{\hat{P}}^c_{M_l}, & d x_{l}^\dagger, & -x_l^\dagger x_l, & \mathbb{I}
 \end{bmatrix}.
 \end{equation}
With these definitions, one can show that $\hat{H} = W^{[1]}\cdot W^{[2]} \hdots \cdot W^{[L-1]} \cdot W^{[L]}$, with $\cdot$ representing a contraction.

This decomposition of $\hat{H}$ offers a highly efficient way to numerically encode the effective Hamiltonian as an MPO.
While we have shown an explicit MPO decomposition for the nearest-neighbor situation, in practice one uses powerful automated packages such as AutoMPO in ITensor~\cite{itensor} to generate compressed forms of the MPO from the second-quantized expressions like Eqs.~\eqref{eq:HA},~\eqref{eq:HB}, and~\eqref{eq:Hdiff}.

\section{TDVP algorithm}\label{app:tdvp}
As described in Sec.~\ref{sec:tn}, we wish to approximate the evolution of the distribution:
\begin{equation}
  \frac{d \ket{p_t}}{dt} = \hat{H} \ket{p_t}
  \label{eq:real_ev}
\end{equation}
by approximating $\ket{p_t}$ as
\begin{align}
  \ket{p_t} &= \sum_{n_1, n_2, \hdots n_L} p_t(n_1,n_2,\ldots,n_L) \ket{n_1, n_2, \hdots n_L} \nonumber\\
    &\approx \hspace{-3mm} \sum_{\substack{n_1,n_2, \hdots,n_L \\ s_1, s_2, \hdots,s_{L-1}}}\hspace{-3mm} Q^{n_1}_{s_1}Q^{n_2}_{s_1s_2}\ldots Q^{n_L}_{s_{L-1}}\ket{n_1, n_2, \hdots n_L} \nonumber\\
&\equiv \ket{q_t}.
\label{eq:Pr_MPS_2}
\end{align}
Here $\ket{q_t}$ is a distribution over microstates that can be expressed as an MPS using the (time-dependent) set of tensors $\left\{Q^{n_1}, Q^{n_2}, \hdots, Q^{n_L}\right\}$.
As in the main text, the distinct $Q$ tensors are implicitly distinguished based on the labeling scheme for the physical index.
Only a subset of possible distributions $\ket{p_t}$ can be written in the form of $\ket{q_t}$, but as the bond dimension $\kappa$ increases, that subset necessarily grows.
We call these distributions which can be represented by MPS states the MPS manifold $\mathcal{M}$.
Because $\hat{H}$ generally evolves $\ket{q_t}$ off $\mathcal{M}$, the Time-Dependent Variational Principle (TDVP)~\cite{dirac1930note,haegeman2011time} seeks the evolution of $\ket{q_t}$ along the MPS manifold:
\begin{equation}
  \frac{d \ket{q_t}}{dt} = \hat{\mathcal{P}}_{T_{\mathcal{M}}} \hat{H} \ket{q_t},
  \label{eq:tdvpeom_2}
\end{equation}
where $\hat{\mathcal{P}}_{T_{\mathcal{M}}}$ is a projector mapping the infinitesimal change in probability $\hat{H} \ket{q_t}$ onto the MPS manifold's nearest (with respect to the $l^2$ norm) tangent vector.
The projector in this equation of motion ensures that $\ket{q_t}$ stays on $\mathcal{M}$ for all times.

Remarkably, in the case of the MPS manifold, an explicit form for the projector can be written down as a sum of $2L-1$ terms, $L$ of which have a correspondence to the $L$ tensors of the MPS and $L-1$ of which have a correspondence to the bonds.
To write down that decomposition, it is beneficial to first define the environment tensors which were obliquely referenced in Sec.~\ref{sec:tn}.
Through gauge transformations, $\ket{q_t}$ is expressed as
\begin{widetext}
  \begin{equation}
    \ket{q_t} = \sum_{\substack{n_1,n_2, \hdots,n_L \\ s_1, s_2, \hdots,s_{L-1}}} A^{n_1}_{s_1} A_{s_1s_2}^{n_2} \hdots A_{s_{l-2}s_{l-1}}^{n_{l-1}} Q_{s_{l-1}s_{l}}^{n_l} B_{s_l s_{l+1}}^{n_{l+1}} \hdots B_{s_{L-2}s_{L-1}}^{n_{L-1}}B_{s_{L-1}}^{n_L} \ket{n_1, \hdots, n_L},
  \label{eq:mixedcanonical_2}
  \end{equation}
\end{widetext}
where the $A$ tensors are left-orthogonal and $B$ tensors are right-orthogonal, meaning
\begin{equation}
\sum_{n_i, s_{i-1}} A_{s_{i-1}s_i}^{n_i} A_{s_{i-1}s_{i}'}^{n_i} = \delta_{s_{i}s_{i}'}
\end{equation}
and
\begin{equation}
\sum_{n_i, s_{i}} B_{s_{i-1}s_i}^{n_i} A_{s_{i-1}'s_{i}}^{n_i} = \delta_{s_{i-1}s_{i-1}'}.
\end{equation}
Summing the $A$ tensors found in Eq.~\eqref{eq:mixedcanonical_2} up to $l-1$ gives the left environment tensor
\begin{align}
  \nonumber & \ket{\Phi_{L, s_{l-1}}^{[n_1:n_{l-1}]}} = \\
  & \ \ \sum_{\substack{n_1,\hdots,n_{l-1} \\ s_1,\hdots,s_{l-2}}} A_{s_1}^{n_1} A_{s_1, s_2}^{n_2} \hdots A_{s_{l-2} s_{l-1}}^{n_{l-1}} \ket{n_1, n_2, \hdots n_{l-1}}
\end{align}
with physical indices counting the occupancy of sites $1$ through $l-1$ and a single uncontracted bond index $s_{l-1}$.
A similar sum over the $s$ indices of the $B$ tensors gives the right environment tensor
\begin{align}
  \nonumber &\ket{\Phi_{R, s_l}^{[n_{l+1}:n_L]}} = \\
  &\sum_{\substack{n_{l+1},\hdots,n_L \\s_{l+1}, \hdots, s_{L-1}}} B_{s_l, s_{l+1}}^{n_{l+1}} \hdots B_{s_{L-2}s_{L-1}}^{n_{L-1}} B_{s_{L-1}}^{n_L} \ket{n_{l+1}, n_{l+2}, \hdots, n_{L}}
\end{align}
with physical indices counting the occupancy of sites $l+1$ through $L$ and a single uncontracted bond index $s_l$.

In terms of the left and right environment tensor projectors,
\begin{equation}
  P_{L}^{[1:l-1]} = \sum_{s_{l-1}} \ket{\Phi_{L, s_{l-1}}^{[n_1:n_{l-1}]}} \bra{\Phi_{L, s_{l-1}}^{[n_1:n_{l-1}]}}
\end{equation}
and
\begin{equation}
  P_{R}^{[l+1:L]} = \sum_{s_{l}} \ket{\Phi_{R, s_{l}}^{[n_{l+1}:n_L]}}\bra{\Phi_{R, s_{l}}^{[n_{l+1}:n_L]}},
\end{equation}
we can now write out the $2L-1$ terms of the projector for the MPS manifold~\cite{haegeman2013post,lubich2015time}:
\begin{align}
\proj_{T_{\mathcal{M}}} &= \sum_{l=1}^L P_L^{[1:l-1]}\otimes \mathbb{I}_l\otimes P_R^{[l+1:L]} - \sum_{l=1}^{L-1} P_L^{[1:l]}\otimes P_R^{[l+1:L]},\nonumber \\
&\equiv \sum_{l=1}^L P_l^+ + \sum_{l=1}^{L-1} P_l^-.
\label{eq:TDVP_proj}
\end{align}
The positive terms $P^+_l$ are associated to the $L$ tensors and the negative terms $P^-_l$ are associated to the $L-1$ bonds.

Integrating Eq.~\eqref{eq:tdvpeom_2} for a time step $\delta t$ gives
\begin{equation}
 \ket{q_{t+\delta t}}= \exp\left(\delta t \left[\sum_{l=1}^L P_l^+ \hat{H} + \sum_{l=1}^{L-1} P_l^- \hat{H}\right]\right) \ket{q_t}
\end{equation}
In the limit of an infinitesimal time step, we can factorize that exponential to act with one $l$ at a time.
For example,
\begin{align}
\nonumber  &\ket{q_{t+\delta t}} = \\
& \ \exp\left(\delta t \left[\sum_{l=2}^L P_l^+ \hat{H} + \sum_{l=2}^{L-1} P_l^- \hat{H}\right]\right) e^{\delta t P_1^-\hat{H}} e^{\delta t P_1^+ \hat{H}} \ket{q_t}.
\label{eq:infinitesimaltime}
\end{align}
The operator $e^{\delta t P_1^+\hat{H}}$ has the physical interpretation of mapping site $1$ forward in time by a time step $\delta t$.
To see this interpretation, we illustrate the action of $e^{\delta t P_2^+ \hat{H}}$ on $\ket{q_t}$, where we assume the MPS has already be written in a mixed canonical form centered on site 2.
The operator $e^{\delta t P_2^+ \hat{H}}$ acts globally on the MPS state, but as long as $\ket{q_t}$ is expressed in the mixed canonical form, only the tensor at site 2 will be altered.
In that way, the $P_2^+$ term is implemented as a \emph{local} update to a single tensor, a local update with the structure of a time evolution with a local effective Hamiltonian.
The reduction from global to local operations is seen if the global exponential operator is expanded in a Taylor series:
\begin{align}
  \nonumber &e^{\delta t P_2^+ \hat{H}} \ket{q_t} \\
  & \ \ \ \ = \sum_{\substack{n_1,n_2, \hdots,n_L \\ s_1, s_2, \hdots,s_{L-1}}} \sum_{k=0}^\infty \frac{\delta t^k}{k!} (P_2^+ \hat{H})^k Q^{n_1}_{s_1}Q^{n_2}_{s_1s_2}\ldots Q^{n_L}_{s_L}\ket{\bfn} \hspace{4in}
\end{align}
Each term of the sum over $\bfn$ can be expressed graphically, where the blue triangular MPS tensors indicate their left/right orthogonality by the direction they point toward the black diamond tensor at the center of the mixed canonical form:
\begin{widetext}
\begin{center}
 \includegraphics{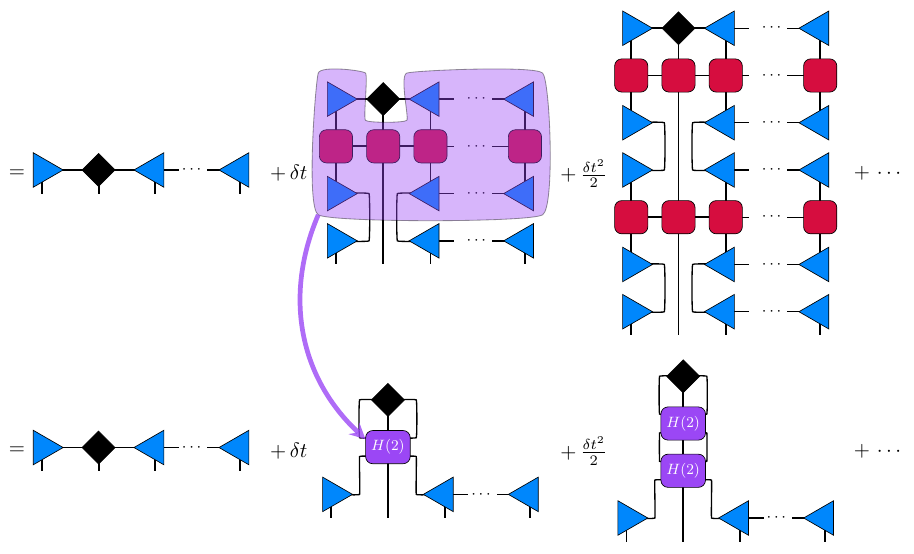}
\end{center}
\end{widetext}
By contracting over all of the tensors in the shaded purple region, we obtain the local tensor we call $H(2)$.
Notice that the same tensor appears repeatedly in the higher order terms of the Taylor series.
These terms can be re-summed to give back a \emph{local} matrix exponential acting on site 2.
Hence, by contracting $Q^{n_2}(t)$ (the black diamond) with $e^{\delta t H(2)}$we get $Q^{n_2}(t+\delta t)$.

Following the $P^+_2 \hat{H}$ propagation, the newly updated tensor $Q^{n_2}(t+\delta t)$ is then decomposed into two blocks $Q^{n_2}_{s_1 s_2} = A^{n_2}_{s_1 s'_2}R_{s'_2s_2}$ via QR or SVD decomposition.
Like how $P^+_2 \hat{H}$ acted locally on the tensor at site 2, $P^-_2$ will act locally to propagate $R_{s'_2s_2}$, but the interpretation is that the propagation is a $\delta t$ time step backwards in time, owing to the negative sign in Eq.~\eqref{eq:TDVP_proj}. $R_{s'_2 s_2}$ is then contracted with the next tensor in the chain, $B_{s_2s_3}^{n_3}$, which is also still at time $t$.
In doing so, we shift gauge such that the new $t + \delta t$ tensor assigned to site 2 is $A^{n_2}_{s_1 s'_2}$ and the new center site tensor at site 3 becomes $Q^{n_3}_{s'_2s_3} = \sum_{s_2}R_{s'_2 s_2} B^{n_3}_{s_2 s_3}$.
Having shifted the center to site 3, one then iterates: acting with $P^+_3 \hat{H}$, performing another QR or SVD decomposition, acting with $P^-_3 \hat{H}$, shifting the gauge, then continuing to sweep down the line.
Sweeping over the entire MPS from $n_1$ to $n_L$ evolves $\ket{q_t}$ by $\delta t$.

\section{Modified Absorbing Boundary Condition Dynamics}\label{app:modifieddynamics}
To measure a rate from $\mathcal{A}$ into $\mathcal{B}$ without multiple crossings of $\mathcal{B}$'s boundary, we define a modified dynamics via an effective Hamiltonian that zeroes out the rates of any reaction or diffusion event that would depart the region $\mathcal{B}$.
Deriving the forms of $\tilde H^{\rm rxn}$ and $\tilde H^{\rm diff}$ in Eq.~\eqref{eq:modified} requires that we first identify the set of states and reactions that depart $\mathcal{B}$:
\begin{equation}
\mathcal{B}^{\rm out} \equiv \left\{\bfn,r\,:\,\bfn + \nu^r \in \mathcal{B}^c\;|\; \bfn \in \mathcal{B} \right\},
\label{eq:Bout}
\end{equation}
where $\mathcal{B}^c$ is the complement of $\mathcal{B}$.
For the Sch\"ogl model, only diffusion and the forward reactions can map out of $\mathcal{B}$.
The states in $\mathcal{B}^{\rm out}$ are those where particles in the $l^{\rm th}$ voxel are at the boundary to $\mathcal{B}$, $n_l = q_\mB^*$ and $n_k \in \mB$ $\forall$ $k\neq l$.
The local boundary for $\mB$ in one voxel can be projected onto using $\mathcal{Q} = \ket{q_\mB^*}\bra{q_\mB^*}$.
Combining $\mathcal{Q}$ with the projector $\mathcal{N}_l = \sum_{n_l\in \mB} \ket{n_l}\bra{n_l}$, lets us define the projection operator 
\begin{equation}
\proj^{\rm out}_l = \mathcal{N}_1\otimes \mathcal{N}_2 \otimes \cdots \mathcal{N}_{l-1} \otimes \mathcal{Q}_l\otimes \mathcal{N}_{l+1}\otimes\dots \mathcal{N}_L
\label{eq:Dl}
\end{equation}
that maps onto states which are at the boundary in voxel $l$ but anywhere in $\mathcal{B}$ in the other voxels.
To build a projector that isolates all boundary configurations, we require that at least one voxel is on the boundary.
We can build such a projector from the powerset expansion
\begin{equation}
\proj^{\rm out} = \sum_{l=1}^L (-1)^{l+1}\sum_{\{s_l\}}\prod_{s_l}\proj^{\rm out}_{s_l},
\label{eq:Pout}
\end{equation}
where $\{s_l\}$ is the set of combinations from $L$ elements taken $l$ at a time.

For $L=3$ Eq. (\ref{eq:Pout}) becomes
\begin{align}
\proj^{\rm out} &= [\proj_1^{\rm out}+\proj_2^{\rm out}+\proj_3^{\rm out}] \nonumber \\
 &-[\proj_1^{\rm out}\proj_2^{\rm out} + \proj_1^{\rm out}\proj_3^{\rm out} + \proj_2^{\rm out}\proj_3^{\rm out}]\nonumber \\
  &+\proj_1^{\rm out}\proj_2^{\rm out}\proj_3^{\rm out}. \nonumber
\end{align}
 Each product over $\proj_{s_l}^{\rm out}$ at multiple voxels can be simplified noting that, $\mathcal{N}_l \mathcal{Q}_l = \mathcal{Q}_l$ which leads to
\begin{equation}
\proj^{\rm out} = \sum_{i=1}^L (-1)^{l+1}\sum_{\{c_k\}} \prod_{c_k} \{\mathcal{N},\mathcal{Q}\}_{c_k},
 \label{eq:DOp}
 \end{equation}
where the sum over $c_k$ is now all combinations of $\mathcal{N}$ and $\mathcal{Q}$ where $\mathcal{Q}$ appears $k$ times. The advantage of Eq. (\ref{eq:DOp}) is that effective Hamiltonians can be projected onto only using local (single-voxel) projectors.
In this work, $\mB$ consists of all molecule numbers less than $q_\mB^* +1$. Then, projecting onto reactions that can map out of $\mB$ the effective Hamiltonians for the Schl\"ogl model become
\begin{align}
\tilde H^{\rm rxn} &= \sum_{l=1}^L \left[\hat H_l^{A_{\rm rev}} + \hat H_l^{B_{\rm rev}} + (\hat H_l^{A_{\rm for}} + \hat H_l^{B_{\rm for}})(\mathbb{I}- \proj^{\rm out})\right], \\
\tilde H^{\rm diff} &= \hat H^{\rm diff}(\mathbb{I}-\proj^{\rm out}).
\end{align}

\section{Error in TDVP rate calculations}
\label{app:Err_kAB}
When calculating observables using the ensemble method, sampling error is avoided, but there are several different sources for numerical error.
Each source of error is controllable in that a parameter can be increased or decreased to systematically improve accuracy at the expense of a greater numerical cost.
That trade-off between accuracy and cost is not straightforward, particularly because it can switch depending on which source of error dominates.
The factorization of the time propagation in Eq.~\eqref{eq:infinitesimaltime} relied upon a infinitesimal time step, but the computational expense grows essentially linearly with the number of timesteps~\footnote{We say essentially linearly rather than linearly because as time progresses the distribution can build richer correlations which can result in singular value decompositions that utilize the full bond dimension $\kappa$. Distributions with less rich correlations can be described by smaller tensors if the singular values are truncated at a threshold.}. 
Consequently, one seeks a time step which is small enough but not too small.
We highlight that the SSA has an effective time step that emerges naturally as the typical time between subsequent events, and that this time step shrinks as the system size grows.
By contrast, the TDVP approach's time step is a parameter one can choose irrespective of the system size.
In addition to the time-step error, the propagation of the local tensors discussed in App~\ref{app:tdvp} is implemented via a Krylov expansion. We found that 30 Krylov basis vectors were sufficient, but the model system here is not very sensitive to the number used.
We have not presented a thorough analysis of time-step or Krylov errors because it is clear that our error is dominated by the bond dimension $\kappa$, which can severely limit the variational search space of allowable MPS distributions.

\begin{figure}[t!]
\centering
\includegraphics[width=.45\textwidth]{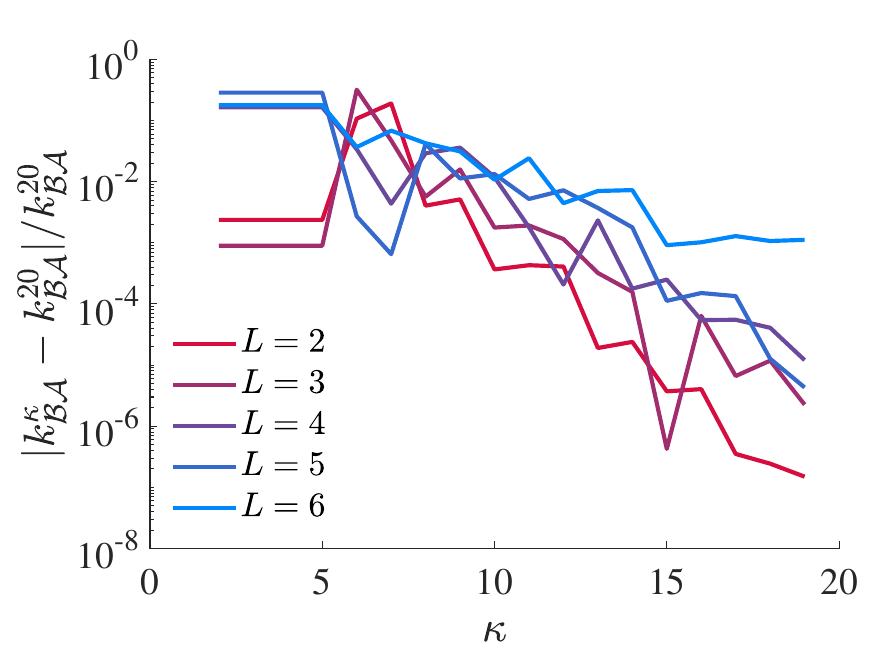}
\caption{\label{fig:Err_VaryBDL} The rate $k_{\mB\mA}$ is calculated at different values of $L$ and different bond-dimensions $\kappa$. The relative difference in rates is plotted against the ``reference" rate calculated at $\kappa = 20$. for large enough bond dimension errors shrink exponentially.}
\end{figure}

In order to get an estimate of how the bond dimension impacts the numerical errors, we repeated the TDVP rate for varying $\kappa$ and $L$.
We would ideally compare each rate against an exact rate, but getting those exact rates is not numerically feasible for all system sizes considered. We have shown (see Fig.~\ref{fig:RateL}c) that for $L = 2$ through $6$, the bond dimension $\kappa = 20$ was sufficient to reproduce the SSA rates to within the size of the plot markers.
For the purposes of analyzing the bond-dimension convergence, we thus take that $\kappa = 20$ calculation as the ``converged'' reference value and measure the approach to that value as $\kappa$ increases from 2 to 20.
Fig.~\ref{fig:Err_VaryBDL} shows the absolute fractional error comparing the rate at a given bond-dimension $\kappa$ and the $\kappa = 20$ reference rate, all with the same time step and Krylov dimension.
We see that errors can already be very small at modest bond dimension (where the smaller tensors make the calculations substantially faster), and the errors shrink exponentially with growing $\kappa$ for all values of $L$.

\section{Runtime Analysis}\label{app:benchmarking}

Fig.~\ref{fig:RateL}c shows the total computational expense to estimate rates from the initial $\mA$ ensemble.
The dominant purpose of communicating the number of CPU hours is to illustrate the scaling of computational expense with the system size.
That is to say, the brute force SSA rates grow exponentially more costly as $L$ increases while the TDVP cost grows subexponentially.
It is enticing to compare the costs to each other.
At a fixed $L$, was the SSA calculation or the TDVP calculation cheaper?
Fig.~\ref{fig:RateL}c appears to answer that question, showing that the two methods have a cross-over with TDVP becoming cheaper for sufficiently large $L$.
We argue that the existence of the cross-over is a generic consequence of the different scalings with $L$, but the exact location of the cross-over depends on specific details of the implementations of both SSA and TDVP.
Considering the difference between the $\kappa = 20$ (black) and $\kappa = 30$ (blue) lines in Fig.~\ref{fig:RateL}c, one must be especially careful to compare one method versus the other.
Each method has parameters that can be adjusted to improve accuracy, and to compare different CPU hour timings, one should have a sense that the two calculations are returning comparable accuracies.
Even more importantly, one should recognize that although we have used the methods to compute the rate, the computational expense goes into two very different objects.
In the case of SSA, that expense results in a collection of representative trajectories, while for TDVP it yields time-dependent joint distributions.
If one were using the calculations to compute different quantities (say waiting time distributions rather than just the rate), the same TDVP joint distributions that were generated to compute the rates could be used at essentially no additional computational expense.
By contrast, generating a converged waiting time distribution from SSA samples would require many more than the 400 trajectories used to estimate the rate.
In a similar vein, in Fig.~\ref{fig:ExpMol} we showed how the TDVP calculations give converged insight into the mechanism of the rare event, but the SSA approach would return only 400 representative events from which one would extrapolate.

\begin{figure}[!h]
\centering
\includegraphics[width=.45\textwidth]{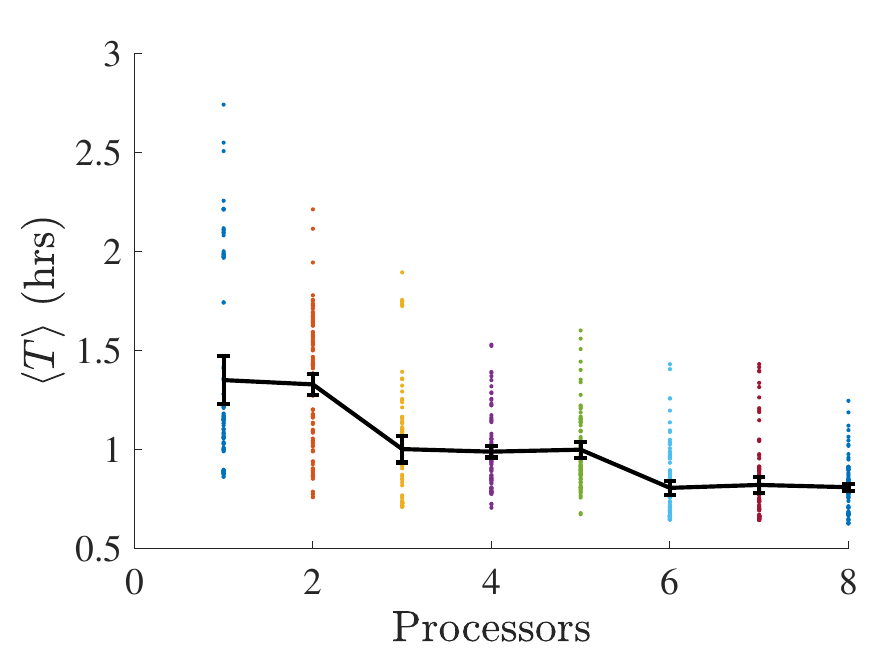}
\caption{\label{fig:BM}
  The wall time to complete 100 time steps of TDVP dynamics with $L = 3$ and $\kappa = 30$ as a function of the number of processors. Parallelization was not explicitly designed into the TDVP sweep, so the limited parallelism emerged only from the built in capabilities of ITensor.
  If one could reach perfect parallelization, either through a clever parallelization scheme or by just running TDVP in serial on a single processor, the number of CPU hours needed to compute the rates in Fig.~\ref{fig:RateL} would decrease by up to a factor of four.}
\end{figure}

Having laid out the caveats, we spell out explicitly how we computed the total computational expense for these two classes of rate calculations (further information about forward flux sampling calculations follows in the next appendix).
The TN calculations are not trivial to parallelize since the TDVP algorithm proceeds from one tensor to the next in serial, with the output of one tensor operation feeding into the next one.
We did not seek a sophisticated parallelization, but we did note some imperfect parallelization could be easily accessed from the way the ITensor software in the Julia language handles the organization and contraction of tensors.
Fig.~\ref{fig:BM} shows that this imperfect parallelization gave a roughly two-fold speed-up if 8 processors were used instead of a single processor.
The computational expense reported in Fig.~\ref{fig:RateL}(c) is the total number of CPU hours for those 8 processors to complete the single-site TDVP dynamics with time step $\delta t = 10^{-4}$ for 10,000 time steps with 30 Krylov dimensions starting from an already computed distribution in $\mA$.
In a sense, this timing is artificially worse than it would need to be.
If we were willing to run for twice the duration of real time, we could decrease the total number of CPU hours for TDVP rates in Fig.~\ref{fig:RateL}(c) by a factor of 4 with a serial implementation.

Whereas the TDVP timing would have room for improvement with more sophisticated parallelization schemes, the SSA calculations are already fully optimized in that they are perfectly parallelizable.
Each stochastic trajectory can and should be run on a separate CPU.
The CPU hours reported in Fig.~\ref{fig:RateL}(c) is thus the sum of the wall times for 400 independent realizations, initialized in $\mA$ and run until $\mB$ was first reached.
The cost for SSA calculations would shift up and down the log scale if one decided to use more or less than the 400 independent trajectories.
Our choice here was to seek a number of trajectories that could make the SSA errorbars on the same order as the plot markers of Fig.~\ref{fig:RateL}(a).

\section{Comparison with Forward Flux Sampling}\label{app:FFS}
\begin{figure}[!h]
\centering
\includegraphics[width=.45\textwidth]{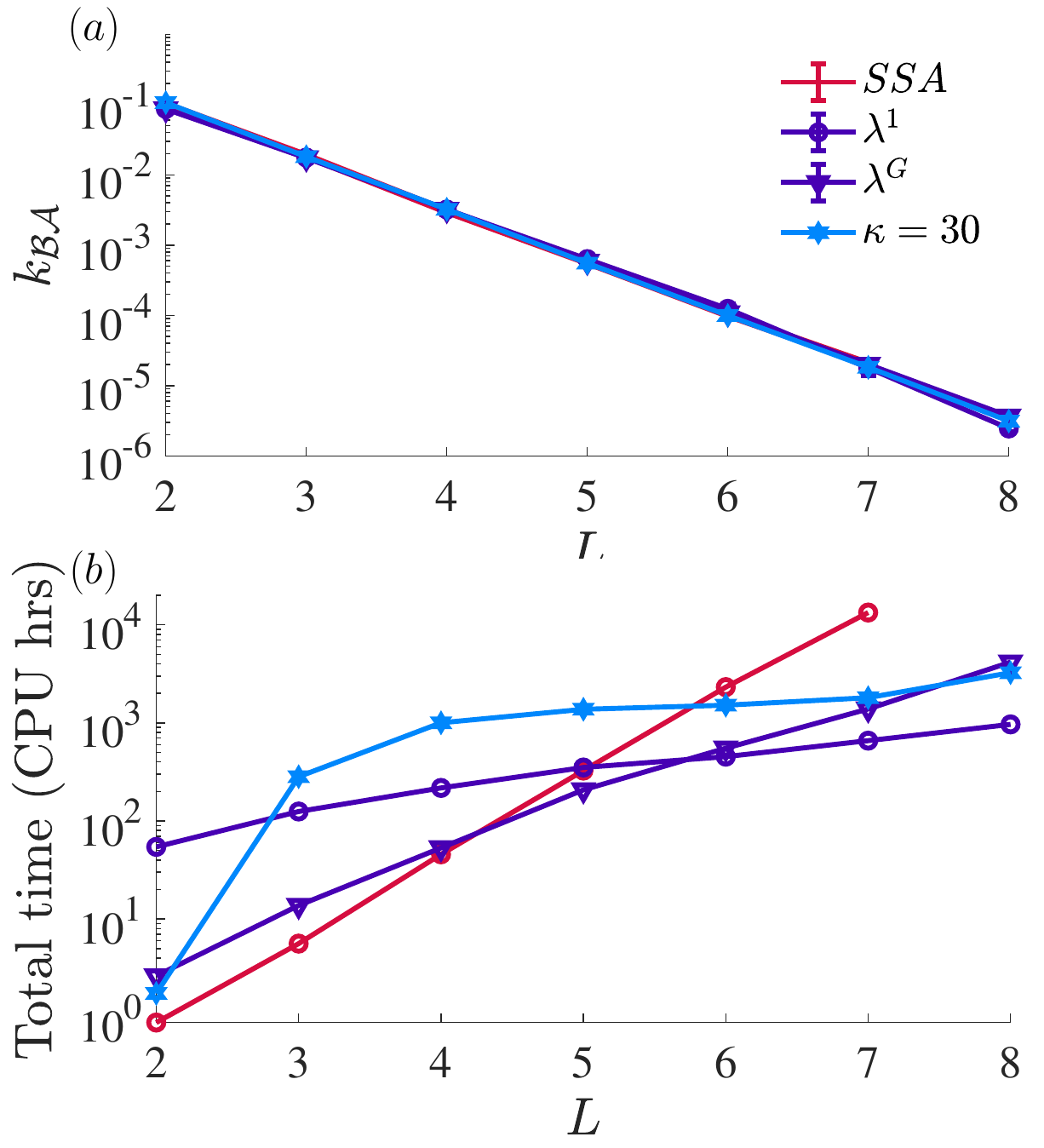}
\caption{\label{fig:FFS_Fig} (a) The rate $k_{\mB\mA}$ from SSA and TDVP using $\kappa = 30$ bond-dimension compared to FFS using different order parameters. Though FFS can correctly estimate the rate even when the order parameter is not exactly the reaction coordinate, the choice of order parameter significantly impacts the computational expense.
  It is notable that the TDVP rate calculation can bypass the introduction of an order parameter; in fact, by analyzing the TDVP-evolved joint distribution, one can reveal the reaction mechanism as an output of the calculation.}
\end{figure}
It is interesting to see how the computational expense of our rate calculation compares to advanced sampling rate calculations.
Here we compare against the original FFS algorithm \cite{allen2005sampling}.
The key idea of FFS is that a (typically one-dimensional) progress coordinate is defined to decompose the single rare event from $\mA$ to $\mB$ down into less rare excursions that make partial progress along that order parameter.
The progress coordinate is discretized to define a collection of $N$ non-overlapping interfaces $\lambda_1, \lambda_2, \hdots, \lambda_N$.
The FFS procedure decomposes the transition rate as
\begin{equation}
k_{\mB\mA} = \Phi_{1,\mA}\prod_{k=1}^{N-1}P(\lambda_{k+1}|\lambda_k),
\end{equation}
where $\Phi_{1, \mA}$ is the flux to pass from $\mA$ to the $\lambda_1$ interface and $P(\lambda_{k+1} | \lambda_k)$ is the probability of reaching $\lambda_{k+1}$ before first returning to $\mA$ conditioned upon initialization at the $\lambda_k$ interface.
To the extent that the chosen progress coordinate acts as a good reaction coordinate, the method can radically improve the efficiency of a rate calculation compared to brute force SSA, but if the order parameter is not ``close enough'' to the reaction coordinate, the conditional probability terms can become very expensive to compute~\cite{hussain2020studying}.
It is tempting to think that $P(\lambda_{k+1}|\lambda_k)$ requires a trajectory starting at $\lambda_k$ and run until it hits either the next or the previous interface, but the actual procedure requires the trajectory to hit either the next interface or $\mA$.
When the progress coordinate poorly approximates the reaction coordinate, a tremendous amount of time can be wasted simulating trajectories slowly wandering back to $\mA$ from one of the final interfaces.
The efficiency of FFS calculations additionally depends on the number of interfaces, their distance from each other, and the number of trajectories needed to resolve the conditional distributions.
Given all the tunable knobs that go into a FFS rate calculation, it only makes sense to compare the numerical expense of TDVP rate calculations against a prudently designed FFS with a reasonable choice of progress coordinate.

Without a priori knowing the mechanism of the Schl\"ogl reaction-diffusion switching events, we designed two different order parameters and performed calculations with both.
For the first order parameter, we focus on the global number of X molecules, summed over all voxels.
The interfaces for this order parameter measuring global progress are $\lambda^G \equiv \left\{\mA, \lambda^G_1, \hdots, \lambda^G_8, \mB\right\}$ with $\lambda^G_i = L \lambda^*_{\mA} + i L$.
Remember that $\lambda^*_{\mA} = 15$ and $\lambda^*_{\mB} = 25$, so this order parameter tracks the progress of the whole system in moving from less than $15L$ molecules to more than $25L$ molecules.
The second order parameter, inspired by Fig.~\ref{fig:ExpMol}'s demonstration that the voxels on the end decrease first, uses only the number of molecules in the first voxel as the sign of progress.
The interfaces for this order parameter measuring voxel 1 are $\lambda^1 \equiv \left\{\mA, \lambda^1_1, \hdots, \lambda^1_8, \mB\right\}$ with each $\lambda^1_i$ interface recording the addition of another X molecule in voxel 1 ($\lambda^1_i =15+i$), so that $\lambda^1_1 = 16$ and $\lambda^1_8 = 23$.

There are many ways to optimize both FFS calculations, for example, changing the number of interfaces, the number of trajectories to estimate $\Phi_{1, \mA}$, and the number of trajectories to estimate the conditional probabilities.
These changes impact the computational expense, and it is not generally knowable how to tune those parameters to most cheaply decrease the errorbars of the rate calculation.
We found that 4,000 trajectories were sufficient to get decent estimates of the flux $\Phi_{1, \mA}$ for both order parameters, though the expense of computing that flux was much greater for $\lambda^G$ than for $\lambda^1$.
While the flux was more costly to compute for $\lambda^G$, the conditional probabilities were easier.
Each conditional probability could be well estimated with only 4,000 additional trajectories per interface.
Converging those conditional probability terms for $\lambda^1$ required $10^7$ trajectories per interface, a cost that was at least partially offset by the cheaper $\Phi_{1, \mA}$ calculation.
For each order parameter we generated these samples to calculate ten independent estimates of the rate, allowing us to measure both the mean and standard error.
From these standard errors, we determined that the FFS calculations were converged in line with the SSA errors in that plot.

A notable feature of FFS is that it will return the correct transition rates even when the order parameter is not a perfect reaction coordinate.
Fig.~\ref{fig:FFS_Fig}a reflects that in that both $\lambda^G$ and $\lambda^1$ reproduce the rates given by the SSA and TDVP methods.
However, the computational expense of the two are not identical.
Fig.~\ref{fig:FFS_Fig}b plots the CPU hours needed to generate the samples that yield the means and standard errors of Fig.~\ref{fig:FFS_Fig}a.
At small $L$, $\lambda^1$ is faster; at large $L$, $\lambda^G$ is faster.
The FFS calculations have many tunable parameters.
They could likely be improved through further refinement of order parameters, changing the number of interfaces, fine-tuning the number of samples for each conditional probability, etc.
Still, we find it notable that our reasonable but perhaps unoptimized FFS implementations perform with similar (and slightly greater) expense as the TDVP rate calculations.
This observation is particularly notable since the TDVP calculations do not need one to guess an order parameter and since they offer converged statistical information about the mechanism in addition to just giving the rate.

\bibliography{references}
\end{document}